\documentstyle[11pt,epsf]{article}
\newcommand{\sect}[1]{ \section{#1} \setcounter{equation}{0} }

\newcommand{\kslash}{k \! \! \! /}

\newcommand{\Dslash}{D \! \! \! \! /}

\newcommand{\half}{\mbox{\small{$\frac{1}{2}$}}}

\newcommand{\Nf}{N_{\!f}} 
\newcommand{\MSbar}{\overline{\mbox{MS}}} 
\begin{document}
\title{Determination of the anomalous dimension of gluonic operators in deep
inelastic scattering at $O(1/N_{\! f})$}  
\author{J.F. Bennett \& J.A. Gracey, \\ Theoretical Physics Division, \\ 
Department of Mathematical Sciences, \\ University of Liverpool, \\ 
Peach Street, \\ Liverpool, \\ L69 7ZF, \\ United Kingdom.}
\date{} 
\maketitle 
\vspace{5cm}
\noindent
{\bf Abstract.} Using large $\Nf$ methods we compute the anomalous dimension of
the predominantly gluonic flavour singlet twist-$2$ composite operator which 
arises in the operator product expansion used in deep inelastic scattering. We 
obtain a $d$-dimensional expression for it which depends on the operator moment
$n$. Its expansion in powers of $\epsilon$ $=$ $(4-d)/2$ agrees with the 
explicit exact three loop $\MSbar$ results available for $n$ $\leq$ $8$ and 
allows us to determine some new information on the explicit $n$-dependence of 
the three and higher order coefficients. In particular the $n$-dependence of 
the three loop anomalous dimension $\gamma_{gg}(a)$ is determined in the 
$C_2(G)$ sector at $O(1/\Nf)$.  

\vspace{-19cm} 
\hspace{13.5cm} 
{\bf LTH-414} 

\newpage

\sect{Introduction.} 
Perturbative quantum chromodynamics, (QCD), is widely accepted as being the 
tool to describe high energy scattering processes involving the strong 
interactions. (For a review see, for example, \cite{1}.) This is primarily due 
to its property of asymptotic freedom which implies that as higher energies are
attained the strength of the strong coupling decreases and perturbation theory 
becomes a more valid approximation. One of the current topics of interest is in
constructing the perturbative series at a higher order than is presently known 
for the anomalous dimensions of the twist-$2$ operators which appear in the 
operator product expansion, which is fundamental to the mathematics underlying 
deep inelastic scattering. One reason that these are required is that they 
arise in the renormalization group equation for the evolution of the 
coefficient functions which are also required at a higher loop order. Indeed 
knowledge of the three loop operator dimensions and two loop coefficient 
functions will mean that the full evolution with energy scale can be performed 
at a new order. The anomalous dimensions have been studied at successive orders
since the discovery that QCD was asymptotically free \cite{2}. The original one
loop calculations were carried out in \cite{3} whilst two loop corrections were
constructed later both for the flavour non-singlet and singlet cases in 
\cite{4,5,6}. These results were expressed not only as a function of the colour
group Casimirs and number of quark flavours, $\Nf$, but also of the moment of 
the operator denoted by $n$. To two loops the expressions as a function of $n$ 
were reasonably straightforward to achieve. One requires this explicit 
dependence rather than, for example, a table of numerical values, in order to 
make contact with an alternative method of calculation. This is the splitting 
function approach of the DGLAP formalism, \cite{7}. The relation between the 
two being determined by the Mellin transform and its inverse where the 
conjugate variable to $n$ is $x$ which corresponds to the momentum fraction 
carried by the parton. Recently substantial progress has been made towards the 
determination of the three loop terms for the operator dimensions. In 
\cite{8,9} exact expressions have been given for the even moments for the 
non-singlet, ($n$ $\leq$ $10$), and singlet, ($n$ $\leq$ $8$), operators. These
calculations were performed by use of computer algebra and symbolic 
manipulation techniques to handle the huge number of Feynman diagrams which 
occur and to organise the resulting expressions for each graph. The 
corresponding coefficient functions are also known for the same values of $n$,
\cite{8,9}. Therefore it remains to construct the full $n$-dependent result. 
Clearly this is a highly non-trivial but important challenge. 

Some insight into the form of the operator dimensions has, however, already 
been provided through the large $\Nf$ technique, \cite{10,11}. This is an 
alternative method to conventional perturbation theory where one can express 
all orders results in the coupling constant of a renormalization group function
as a closed function of the spacetime dimension $d$. For QCD the method has 
extended the pioneering work for the $O(N)$ $\sigma$ model developed in the 
series of articles \cite{12,13}. Through ideas in the critical renormalization 
group one can directly relate the coefficients in the $\epsilon$-expansion of 
this function or critical exponent, where $d$ $=$ $4$ $-$ $2\epsilon$, to the 
explicit perturbative coefficients at the same order in powers of $1/\Nf$.
So, for example, using this approach an expression was derived for the 
$\beta$-function of QCD, \cite{14}, and the non-singlet twist-$2$ operator 
dimension at $O(1/\Nf)$, \cite{10}. The three loop coefficient of the latter, 
which was given explicitly as a function of $n$, agrees exactly with the full 
calculation of \cite{8} for the even moments $2$ $\leq$ $n$ $\leq$ $10$. 
Likewise the extension of that work to the singlet case, \cite{11}, yielded 
similar exact agreement for the region of overlap, \cite{9}. Clearly these 
analytic results will play an important role for checking the full 
$n$-dependent three loop result. However, the singlet $1/\Nf$ calculation of 
\cite{11} only concentrated on one of the operators which occurs. In 
perturbation theory there is mixing between two operators which are 
respectively predominantly fermionic and gluonic in nature, \cite{3}. As the 
former was studied in \cite{11} it remains to derive results for the latter 
type of operator in order to complete the full leading order in $1/\Nf$ 
computation. This is the main aim of this paper where we will discuss the 
technical issues required to obtain a result on a par with \cite{11}. Although 
these are provided with the particular case of QCD in mind, the calculational 
algorithms to determine the values of the scalar integrals which arise are 
reasonably general and therefore applicable to the determination of the 
anomalous dimensions of similar singlet operators in other field theories in 
the $1/\Nf$ method. Finally, we remark that although this paper is concerned 
with large $\Nf$ methods for the operator dimensions, some insight into the 
$n$-dependence of the coefficient functions in large $\Nf$ have already been 
provided in \cite{15,16}.   

The paper is organised as follows. We review the basic formalism and 
background to our problem in section 2. The details of the computation of the
Feynman diagrams which occur in the large $\Nf$ critical point formalism are
provided in sections 3 and 4. These deal respectively with the QED and 
non-QED sectors of the operator dimension defined essentially by those graphs
which do not and do vanish by Furry's theorem. Section 5 is devoted to the 
derivation of explicit $n$-dependent results as well as several concluding 
remarks. The appendices respectively provide some basic integral results and
an explicit three loop calculation to illustrate some of the internal cross
checks which were necessary to validate the general recurrence relations we 
had to derive. 

\sect{Background.} 

We begin by reviewing the necessary features of the perturbative structure of
the anomalous dimensions of the twist-$2$ singlet operators we are interested
in. These are defined to be, \cite{3},  
\begin{eqnarray}
{\cal O}^{\mu_1 \ldots \mu_n}_{\mbox{\footnotesize{$q$}}} &=&  
i^{n-1} {\cal S} \bar{\psi}^I \gamma^{\mu_1} D^{\mu_2} \ldots D^{\mu_n} 
\psi^J - \mbox{trace terms} \nonumber \\ 
{\cal O}^{\mu_1 \ldots \mu_n}_{\mbox{\footnotesize{$g$}}} &=&  
\half i^{n-2} {\cal S} \, \mbox{tr} \, G^{a \, \mu_1\nu} D^{\mu_2} 
\ldots D^{\mu_{n-1}} G^{a \,\, ~ \mu_n}_{~~\nu} - \mbox{trace terms}
\label{op1}
\end{eqnarray}
where $\psi^{iI}$ is the quark field, $A^a_\mu$ is the gluon field, $D_\mu$ $=$
$\partial_\mu$ $+$ $T^a A^a_\mu$, $G^a_{\mu\nu}$ $=$ $\partial_\mu A^a_\nu$ $-$ 
$\partial_\nu A^a_\mu$ $+$ $f^{abc} A^b_\mu A^c_\nu$, $T^a$ are the colour 
group generators, $f^{abc}$ are the structure constants, $1$ $\leq$ $i$ $\leq$ 
$N_c$, $1$ $\leq$ $I$ $\leq$ $\Nf$, $1$ $\leq$ $a$ $\leq$ $(N^2_c$ $-$ $1)$ and
${\cal S}$ denotes symmetrization in the Lorentz indices. In perturbation 
theory the canonical dimension of each of the operators is equivalent. 
Therefore they mix under renormalization which leads to a matrix of 
renormalization constants and associated matrix of anomalous dimensions. To fix
notation we define the renormalization constants by  
\begin{equation} 
{\cal O}^{\footnotesize{\mbox{ren}}}_i ~=~ Z_{ij} 
{\cal O}^{\footnotesize{\mbox{bare}}}_j 
\end{equation} 
where the anomalous dimensions, $\gamma_{ij}(a)$, are defined by 
\begin{equation} 
\gamma_{ij}(a) ~=~ \left( 
\begin{array}{ll} 
\gamma_{qq}(a) & \gamma_{gq}(a) \\ 
\gamma_{qg}(a) & \gamma_{gg}(a) \\ 
\end{array} 
\right) 
\label{mixmat} 
\end{equation} 
with $\gamma_{ij}(a)$ $=$ $\beta(a) (\partial/\partial a) \ln Z_{ij}$ and 
$\beta(a)$ is the usual renormalization group function governing the running
of the coupling constant $a$. The entries in $\gamma_{ij}(a)$ depend on the
colour group parameters, $\Nf$ and $n$ and since it is the $1/\Nf$ corrections 
which we will focus on we define the explicit form of each the entries as 
\begin{eqnarray}  
\gamma_{qq}(a) &=& a_1a + (a_{21}\Nf + a_{22})a^2 + (a_{31}\Nf^2 + a_{32}\Nf 
+ a_{33})a^3 + O(a^4) \nonumber \\ 
\gamma_{gq}(a) &=& b_1a + (b_{21}\Nf + b_{22})a^2 + (b_{31}\Nf^2 + b_{32}\Nf 
+ b_{33})a^3 + O(a^4) \nonumber \\ 
\gamma_{qg}(a) &=& c_1 \Nf a + c_2\Nf a^2 + (c_{31}\Nf^2 + c_{32}\Nf 
+ c_{33})a^3 + O(a^4) \nonumber \\ 
\gamma_{gg}(a) &=& (d_{11}\Nf + d_{12})a + (d_{21}\Nf + d_{22})a^2 
+ (d_{31}\Nf^2 + d_{32}\Nf + d_{33})a^3 + O(a^4)  
\label{matdef} 
\end{eqnarray} 
Clearly the $\Nf$ dependence does not enter in the same fashion in each term.
The explicit determination was carried out in \cite{4,5,6} but we record for 
completeness only those which we require
\begin{eqnarray} 
a_1 &=& C_2(R) \left[ 4 S_1(n) ~-~ 3 ~-~ \frac{2}{n(n+1)} \right] ~~~,~~~  
b_1 ~=~ -~ \frac{2(n^2+n+2)C_2(R)}{n(n^2-1)} \nonumber \\
c_1 &=& -~ \frac{4(n^2+n+2)T(R)}{n(n+1)(n+2)} ~~~,~~~  
d_{11} ~=~ \frac{4}{3} T(R) \nonumber \\ 
d_{12} &=& C_2(G) \left( 4S_1(n) - \frac{11}{3} - \frac{4}{n(n-1)} 
- \frac{4}{(n+1)(n+2)} \right) \nonumber \\  
a_{21} &=& T(R)C_2(R) \left[ \frac{2}{3} ~-~ \frac{80}{9}S_1(n) ~+~ 
\frac{16}{3}S_2(n) \right. \nonumber \\ 
&&+~ \left. \frac{8[11n^7+49n^6+5n^5-329n^4-514n^3-350n^2-240n-72]} 
{9n^3(n+1)^3(n+2)^2(n-1)} \right] \nonumber \\ 
b_{21} &=& -~ \frac{16C_2(R)T(R)}{3} \left[ \frac{1}{(n+1)^2} ~+~ 
\frac{(n^2+n+2)}{n(n^2-1)} \left( S_1(n) ~-~ \frac{8}{3} \right) \right] 
\nonumber \\ 
d_{21} &=& C_2(R) T(R) \left( 4 
+ \frac{8(2n^6+4n^5+n^4-10n^3-5n^2-4n-4)}{(n-1)n^3(n+1)^3(n+2)} \right)
\nonumber \\ 
&& +~ C_2(G) T(R) \left( - \frac{80}{9}S_1(n) + \frac{16}{3}  
+ \frac{8(38n^4+76n^3+94n^2+56n+12)}{9(n-1)n^2(n+1)^2(n+2)} \right) 
\label{pertcoeff} 
\end{eqnarray} 
where the colour group Casimirs are defined by $T^a T^a$ $=$ $C_2(R)$, 
$\mbox{tr}(T^a T^b)$ $=$ $T(R)\delta^{ab}$ and $f^{acd} f^{bcd}$ $=$ $C_2(G)
\delta^{ab}$. The function $S_l(n)$ is defined by $S_l(n)$ $=$ $\sum_{r=1}^n 
1/r^l$. These expressions (\ref{pertcoeff}) are determined in perturbation 
theory by inserting both operators, (\ref{op1}), in quark and gluon 
$2$-point functions and deducing the pole with respect to the dimensional 
regulator before using $\MSbar$ to define $Z_{ij}$. The procedure to find 
leading order $1/\Nf$ information on $\gamma_{ij}(a)$ is similar and has been 
discussed in \cite{10,11}. Briefly one uses critical point renormalization 
group techniques to analyse the scaling behaviour of the operators we are 
interested in at a fixed point in the $\beta$-function of QCD. This is defined 
to be a non-trivial zero of the $d$-dimensional $\beta$-function and the value 
of the coupling is denoted by $a_c$. As has been observed in \cite{14} the 
propagators take a particularly simple scaling form at $a_c$. Further, the 
anomalous dimension of each operator can be deduced in powers of $1/\Nf$ by 
noting that there is a relationship between the critical exponent of the 
operator and the analogous anomalous dimension renormalization group function 
evaluated at $a_c$. In the $d$-dimensional case such a non-trivial zero exists 
and, moreover, is computable in a power series in $1/\Nf$. For instance, from 
the $3$-loop result, \cite{17,18}, in the notation of \cite{19},  
\begin{eqnarray}
\beta(a) &=& \frac{(d-4)}{2}a + \left[ \frac{4}{3}T(R)\Nf - \frac{11}{3}C_2(G) 
\right] a^2 \nonumber \\
&& +~ \left[ 4C_2(R)T(R)\Nf + \frac{20}{3}C_2(G)T(R)\Nf
- \frac{34}{3}C^2_2(G) \right] a^3 + O(a^4) 
\end{eqnarray} 
one deduces that as $\Nf$ $\rightarrow$ $\infty$, omitting terms 
$O(\epsilon^5/\Nf^3)$,  
\begin{eqnarray}
a_c &=& \frac{3\epsilon}{4T(R)\Nf} ~+~ \left[ \frac{33}{16}C_2(G) \epsilon 
{} ~-~ \left( \frac{27}{16}C_2(R) + \frac{45}{16}C_2(G) \right) \epsilon^2 
\right. \nonumber \\ 
&& \left. +~ \left( \frac{99}{64}C_2(R) + \frac{237}{128}C_2(G) \right) 
\epsilon^3 ~+~ \left( \frac{77}{64}C_2(R) + \frac{53}{128}C_2(G) \right) 
\epsilon^4 \right] \frac{1}{T^2(R)\Nf^2}  
\label{ac} 
\end{eqnarray}
where we have included the contribution from the recently computed exact
$4$-loop $\MSbar$ $\beta$-function, \cite{19}.  

The scaling behaviour of the quark and gluon propagators in momentum space at 
$a_c$ are then simply, \cite{10},  
\begin{equation} 
\psi(k) ~ \sim ~ \frac{A\kslash}{(k^2)^{\mu-\alpha}} ~~~,~~~ 
A_{\mu\nu}(k) ~ \sim ~ \frac{B}{(k^2)^{\mu-\beta}} \left[ \eta_{\mu\nu} ~-~ 
(1-b) \frac{k_\mu k_\nu}{k^2} \right] 
\label{props} 
\end{equation}  
where $b$ is the covariant gauge parameter, $d$ $=$ $2\mu$ and the exponents 
$\alpha$ and $\beta$ are defined as  
\begin{equation} 
\alpha ~=~ \mu ~-~ 1 ~+~ \half \eta ~~~,~~~ \beta ~=~ 1 ~-~ \eta ~-~ \chi 
\label{defexp} 
\end{equation} 
The quantity $\eta$ is the quark anomalous dimension which corresponds to the 
quark wave function renormalization at the critical coupling and $\chi$ is the 
anomalous dimension of the quark gluon vertex. The canonical dimensions are
deduced by demanding that the $d$-dimensional action is dimensionless. This is 
contrary to perturbation theory where the canonical part is defined by ensuring
the $4$-dimensional action is dimensionless. The quantities $A$ and $B$ are 
momentum independent amplitudes. Using (\ref{props}) various leading order 
results have been established such as the QCD $\beta$-function at $O(1/\Nf)$, 
\cite{14}. This encodes the coefficients of all powers of $\epsilon$ in the 
$O(1/\Nf^2)$ term of (\ref{ac}). As the leading order values of $\eta$, $\chi$ 
and the amplitude combination $z$ $=$ $A^2B$ are required here, we note their 
explicit values are, \cite{20},  
\begin{eqnarray} 
\eta_1 &=& \frac{[(2\mu-1)(\mu-2)+\mu b] C_2(R)\eta^{\mbox{o}}_1} 
{(2\mu-1)(\mu-2)T(R)} \\  
\chi_1 &=& -~ \frac{[(2\mu-1)(\mu-2)+\mu b] C_2(R)\eta^{\mbox{o}}_1} 
{(2\mu-1)(\mu-2)T(R)} ~-~ \frac{[(2\mu-1)+ b(\mu-1)] C_2(G)\eta^{\mbox{o}}_1} 
{2(2\mu-1)(\mu-2)T(R)} \\  
z_1 &=& \frac{\Gamma(\mu+1)\eta^{\mbox{o}}_1}{2(2\mu-1)(\mu-2)T(R)} 
\end{eqnarray}  
where, for example, $\eta$ $=$ $\sum_{i=1}^\infty \eta_i(\epsilon)/\Nf^i$ and 
\begin{equation} 
\eta^{\mbox{o}}_1 ~=~ 
\frac{(2\mu-1)(\mu-2)\Gamma(2\mu)}{4\Gamma^2(\mu)\Gamma(\mu+1)\Gamma(2-\mu)} 
\end{equation}  

In \cite{11} it was these propagators which were used to analyse the fermionic 
operator ${\cal O}_q$ and produce an expression which agreed with known results
to three loops. This remark needs to be qualified, however, since the canonical
dimensions of the fields at $a_c$ differ from the perturbative ones. Therefore 
the canonical dimensions of (\ref{op1}) are different and so they cannot mix 
under renormalization. In other words the mixing matrix at criticality is 
triangular at least at leading order in $1/\Nf$ and both operators cease being 
twist-$2$ at $a_c$. Instead ${\cal O}_g$ is twist-$(2+O(\epsilon))$. Although 
this would appear to mean that the critical point large $\Nf$ method fails to 
have any contact with perturbation theory which we are seeking, there is a 
relation between $\gamma_{ij}(a_c)$ and the scaling dimensions of (\ref{op1}) 
at $a_c$. This was established in \cite{11} where it was noted that the 
eigen-anomalous dimensions of (\ref{mixmat}) evaluated at $a_c$ are in fact the
dimensions of the operators (\ref{op1}) whose fields have the canonical 
dimensions given by (\ref{defexp}). In other words in perturbation theory the 
information contained in the critical exponents relating to the dimensions of 
(\ref{op1}) at $a_c$ are given by two combinations of the perturbative 
coefficients listed in (\ref{pertcoeff}). To make these comments more 
transparent we return to $\gamma_{ij}(a)$ and deduce the eigen-anomalous 
dimensions at $a_c$ in powers of $1/\Nf$. First, the eigenvalues for general 
$a$ are given by 
\begin{equation} 
\lambda_\pm(a) ~=~ \frac{1}{2} ( \gamma_{qq} + \gamma_{gg} ) ~\pm~ 
\frac{1}{2} \left[ ( \gamma_{qq} - \gamma_{gg} )^2 + 4 \gamma_{qg}\gamma_{gq} 
\right]^{\half} 
\end{equation}  
However, expanding in powers of $a$ and retaining the same orders in $1/\Nf$
with the definitions (\ref{matdef}), we find  
\begin{eqnarray} 
\lambda_-(a) &=& \left( a_1 - \frac{b_1c_1}{d_{11}}\right) a + 
\left( a_{21} - \frac{b_{21}c_1}{d_{11}}\right) \Nf  a^2 + 
\left( a_{31} - \frac{b_{31}c_1}{d_{11}}\right) \Nf^2 a^3 + O(\Nf^3 a^4) 
\nonumber \\ 
\lambda_+(a) &=& \left( d_{11}\Nf + d_{12} + \frac{b_1c_1}{d_{11}} \right) a
+ \left( d_{21} + \frac{b_{21}c_1}{d_{11}} \right) \Nf a^2 \nonumber \\ 
&& ~ + \left( d_{31} + \frac{b_{31}c_1}{d_{11}} \right) \Nf^2 a^3 
+ O(\Nf^3 a^4) 
\end{eqnarray} 
Noting that $a_c$ is $O(1/\Nf)$, it is easy to observe that $\lambda_-(a_c)$ 
$=$ $O(1/\Nf)$ whereas $\lambda_+(a_c)$ $=$ $O(\epsilon)$ which bears out our
remarks on the difference in twist at $a_c$. As the dimension of the 
predominantly fermionic operator has been discussed elsewhere, we record that
the predominantly gluonic eigen-operator dimension is  
\begin{eqnarray} 
\lambda_+(a_c) &=& \frac{3 d_{11} \epsilon}{4} ~+~ \frac{1}{T(R)\Nf} \left[ 
\frac{3}{4} \left( d_{12} + \frac{b_1c_1}{d_{11}} + \frac{11}{4}C_2(G) d_{11} 
\right) \epsilon \right. \nonumber \\ 
&&+~ \left. \frac{9}{16} \left( d_{21} + \frac{b_{21}c_1}{d_{11}}
- \left( 3C_2(R) + 5C_2(G) \right) d_{11} \right) \epsilon^2 \right. 
\nonumber \\ 
&&+~ \left. \frac{27}{64} \left( d_{31} + \frac{b_{31}c_1}{d_{11}}
+ \left( \frac{11}{3}C_2(R) + \frac{79}{18} C_2(G) \right) d_{11} \right) 
\epsilon^3 \right] ~+~ O \left( \frac{\epsilon^4}{\Nf^2} \right) 
\label{eigandim} 
\end{eqnarray} 
We note that the location of $a_c$ is present in this expression because of a
non-zero $d_{11}$ term at one loop in (\ref{matdef}).  

Two technical issues remain to be discussed which are fundamental to the 
calculation of $\lambda_+(a_c)$. The first relates to a property of the fixed
point. At $a_c$ it was demonstrated by Hasenfratz and Hasenfratz, \cite{21}, 
that at $O(1/\Nf)$ QCD is equivalent to a simpler lower dimensional model which
is the non-abelian Thirring model, (NATM). This means that they lie in the same 
universality class and therefore critical exponents are the same in each model.
In other words it does not matter which theory we study at $a_c$, the same 
$d$-dimensional expressions are obtained at leading order. Similar and more
widely studied $d$-dimensional equivalences exist like, for example, that 
between the $O(N)$ $\sigma$ model and $O(N)$ $\phi^4$ theory. (See, for 
example, \cite{22}.) For completeness the respective lagrangians are  
\begin{eqnarray} 
L^{\mbox{\footnotesize{QCD}}} &=& i \bar{\psi}^{iI} ( \Dslash \psi)^{iI} ~-~ 
\frac{(G^a_{\mu\nu})^2}{4e^2} \\ 
L^{\mbox{\footnotesize{NATM}}} &=& i \bar{\psi}^{iI} ( \Dslash \psi)^{iI} ~-~ 
\frac{(A^a_\mu)^2}{2\lambda^2}
\end{eqnarray} 
where $e$ and $\lambda$ are the coupling constants which are dimensionless in 
$4$ and $2$ dimensions respectively. However, since the NATM has fewer 
interactions due to the absence of cubic and quartic gluon self-interactions it
is more practical to use it. The effects of such interactions are recovered in 
diagrams with quark loops with respectively three and four gluon legs, 
\cite{21}. Indeed this equivalence has already proved useful in establishing 
several results in large $\Nf$ QCD, \cite{14}. 

The second point concerns the regularization of the theory at the fixed point.
Unlike perturbation theory we use a fixed non-regularizing dimension $d$ in the
critical point approach and therefore the only sensible type of regulator which
remains to us is an analytic one. This is introduced by shifting the gluon 
dimension $\beta$ to $\beta$ $-$ $\Delta$ where $\Delta$ is a dimensionless 
infinitesimal parameter, \cite{13}. It plays a similar role to the 
$\epsilon$-regularizing parameter of dimensional regularization. Therefore when
we insert ${\cal O}_g$ in a gluon $2$-point function and replace the lines by 
the critical propagators with a non-zero $\Delta$ the value of the diagram is 
no longer infinite but proportional to $1/\Delta$. As has been discussed in 
\cite{13,11} the residue of this simple pole is an $n$ and $d$-dependent 
function and it is this which contributes to the full value of 
$\lambda_+(a_c)$. 

\sect{QED sector.} 

We now turn to the discussion of the technical issues involved in computing
the critical exponent $\lambda_+(a_c)$ beginning with some general 
considerations. As in a perturbative calculation we insert the composite 
operator ${\cal O}_g$ into a gluonic $2$-point Green's function and compute 
its divergent part with respect to the critical regularization. This is 
illustrated in fig. 1 and the total value of such contributions correspond to
the renormalization of the bare operator whose critical exponent is  
$\eta_{{\cal O}_g}$. To obtain the full value one must also include the wave 
function renormalization of the fields present in the composite operator as 
in perturbation theory. Since these are gluons then the dimension is  
\begin{equation} 
\lambda_+(a_c) ~=~ -~ \eta ~-~ \chi ~+~ \eta_{{\cal O}_g}  
\label{expop} 
\end{equation} 
It is worth noting that because the operators (\ref{op1}) are physical the 
combination (\ref{expop}) is independent of the gauge choice even though each
of the individual quantities of (\ref{expop}) depend on $b$. Therefore we 
have chosen to perform the calculation in the Feynman gauge. From \cite{2,4} 
and (\ref{op1}) the Feynman rule for the (bare) insertion which is quadratic
in the gluon field, is, for an external momentum $p$,  
\begin{equation} 
\half [1 + (-1)^n ] \, \left[ \eta_{\mu\nu} (\Delta p)^2 
- (\Delta_\mu p_\nu + \Delta_\nu p_\mu) \Delta p 
+ p^2 \Delta_\mu \Delta_\nu \right] (\Delta p)^{n-2} 
\label{oprule} 
\end{equation}  
where $\mu$ and $\nu$ are the free indices which remain after contracting with 
the null vector $\Delta_\mu$ of the light cone projection, \cite{11}. (Although
we have used the same symbol for the analytic regularization there ought not to
be any confusion in what follows as one is clearly a scalar object whilst the 
other is a Lorentz vector with the property $\Delta^2$ $=$ $0$.) The Feynman 
rule for the insertion with three or more gluon legs have been given elsewhere,
\cite{4}. At the order in $1/\Nf$ which we are considering there are only a few
contributing graphs whch are illustrated in fig. 2. There are no graphs with 
self energy insertions on internal lines since we are computing with 
propagators which have non-zero anomalous dimensions. Each of these graphs in 
fig. 2 will in principle have the same general Lorentz structure in its pole 
part which is  
\begin{equation} 
\Gamma_{\mu\nu} ~=~ \left[  A \eta_{\mu\nu} (\Delta p)^2 
+ B (\Delta_\mu p_\nu + \Delta_\nu p_\mu) \Delta p 
+ C \frac{p_\mu p_\nu}{p^2} (\Delta p)^2  
+ D p^2 \Delta_\mu \Delta_\nu \right] (\Delta p)^{n-2} 
\label{opdeco}
\end{equation}  
where $A$, $B$, $C$ and $D$ will depend on $\mu$, $n$ and the regulator though
it is only the residue of the simple pole in the latter which we are interested
in. In principle we ought to compute each of these expressions for each of the 
integrals, which would require taking various projections on the explicit 
form of each graph. However we can minimize the amount of computation that
needs to be performed by appealing to the renormalizability of QCD. As the 
theory is renormalizable then the sum of the contribution of the form 
(\ref{opdeco}) from each of the graphs of fig. 2 must be proportional to 
(\ref{oprule}). The resulting coefficient of proportionality is then simply
related to $\eta_{{\cal O}_g}$. Therefore it is much simpler to compute only
one of the four quantities in (\ref{opdeco}). (However, we note that we have
checked this assumption by explicit computation for the QED sector of fig. 2
which involves only the first two $2$-loop graphs.) We have chosen to deduce
$A$ and it is a simple exercise to invert (\ref{opdeco}) to produce 
\begin{equation} 
A ~=~ \frac{1}{2(\mu-1)} \left[ \eta_{\mu\nu} (\Delta p)^2 
- (\Delta_\mu p_\nu + \Delta_\nu p_\mu) \Delta p 
+ p^2 \Delta_\mu \Delta_\nu \right] (\Delta p)^{n-2} \Gamma_{\mu\nu}  
\label{ainv} 
\end{equation}  
This projection on each of the four graphs can be simplified further by 
noting that renormalizability also requires that $C$ is finite with respect 
to the regularization. In the same inversion which produces (\ref{ainv}),  
\begin{equation} 
C ~=~ \frac{\Delta_\mu \Delta_\nu}{2(\mu-1)} \Gamma_{\mu\nu} 
\end{equation} 
Therefore the projection of each of the integrals we take to produce the 
divergent part with respect to the regularization is 
\begin{equation} 
\frac{1}{2(\mu-1)} \left[ \eta_{\mu\nu} (\Delta p)^2 
- (\Delta_\mu p_\nu + \Delta_\nu p_\mu) \Delta p 
\right] (\Delta p)^{n-2} \Gamma_{\mu\nu}  
\label{proj} 
\end{equation}  
This step substantially reduces the number of constituent integrals which would
have to be considered in, for example, the final graph of fig. 2. A second 
simplification of the organization of the calculation is to examine the
contributions from the sectors defined by the colour group Casimirs $C_2(R)$ 
and $C_2(G)$ separately. As the latter arises due to non-zero colour group 
structure constants we will refer to the former as the QED sector and 
concentrate on determining its contribution first. One reason for this is that
there are various technical issues in the full calculation which we need to 
discuss and these are best illustrated in this case. In the $C_2(G)$ sector 
other techniques will be introduced which will assume these results. 

For the remainder of the section we will discuss the first two $2$-loop graphs
of fig. 2 as only they involve $C_2(R)$. The first of these graphs is 
elementary to determine, primarily because it is equivalent to chain integrals 
in the language of \cite{12}. However the presence now of an operator with a 
non-zero moment means that more general chain results are required. These have 
been given in appendix A as they have been widely used by other authors in this
area, \cite{23}. Therefore we merely record the total contribution from this 
graph to $\eta_{{\cal O}_g}$ is  
\begin{eqnarray} 
&& [ \mu^2 n^2 + \mu^2 n - 2\mu^2 - 9\mu n^2 - 9\mu n + 2\mu 
- n^4 - 2n^3 \nonumber \\ 
&&~ + 9n^2 + 10n]\mu(\mu-1) \Gamma(\mu)\Gamma(n+2-\mu) C_2(R) \eta_1^{\mbox{o}} 
\nonumber \\ 
&&~~ /[n(n+1)(\mu-2)(\mu-n-1)\Gamma(3-\mu) \Gamma(\mu+n+1) T(R)] 
\end{eqnarray} 

The computation of the second integral of fig. 2 is more tedious to determine. 
From (\ref{proj}) we treat the contributions from the $\eta^{\mu\nu}$ and 
$(\Delta^\mu p^\nu$ $+$ $\Delta^\nu p^\mu)$ projections separately. To proceed 
one contracts with either projection tensor and then takes the trace over the 
quark loop which leaves a set of two loop scalar integrals. The majority of 
these integrals can be calculated in the same way as those of the first graph 
of fig. 1 since they are elementary chains. However a set of two loop graphs 
remain which have the form illustrated in fig. 3 which clearly require another 
technique. We use a coordinate space notation for integrals, similar to 
\cite{12}, but the explicit form of that graph is given by  
\begin{equation} 
G_{pqn}(\alpha) ~=~ \int_{yz} \, \frac{ (\Delta y)^p (\Delta z)^q 
(\Delta(y-z))^n}{y^2 (x-y)^2 z^2 (x-z)^2 ((y-z)^2)^\alpha} 
\label{gdefn} 
\end{equation} 
where $p$, $q$ and $n$ are integers and $\alpha$ is an arbitrary exponent. Here
and elsewhere we will use the notation that a line in a graph which has a 
bracketed integer, $(p)$, beside it corresponds to a factor in the numerator of
$p$ contractions of its vector with $\Delta_\mu$. Unbracketed symbols beside a 
line correspond to its propagator exponent. From the nature of the original 
integral the central $(y-z)$ line corresponds to the location of the operator 
insertion with moment $n$. It turns out that only several values for $p$ and 
$q$ are needed for $n$ arbitrary. These are $(p,q)$ $=$ $(0,0)$, $(1,0)$, 
$(1,1)$ and $(2,0)$. However, some of these are related by symmetry or by 
rewriting $\Delta y$ as $\Delta y$ $=$ $\Delta(y-z)$ $+$ $\Delta z$. Therefore 
for $n$ even 
\begin{equation} 
G_{10n}(\alpha) ~=~ \half G_{00n}(\alpha) 
\end{equation} 
and for arbitrary $n$ 
\begin{equation} 
G_{11n} ~=~ G_{20n}(\alpha) ~-~ \half G_{00 (n+2)}(\alpha) 
\end{equation} 
It therefore remains to compute $G_{00n}(\alpha)$ and $G_{20n}(\alpha)$ 
which is achieved by establishing a recurrence relation for the more general
integrals where the $\Delta^\mu$ contractions of all numerator vectors has not 
been performed. Then we will solve this for the cases we are interested in.  
It is best to illustrate this for the simpler $G_{00n}(\alpha)$ case first.
Then using Lorentz symmetry that integral will decompose into the form 
\begin{equation} 
\left[ A(\alpha,n) x_{\mu_1} \ldots x_{\mu_n} ~+~ 
B(\alpha,n) [ \eta_{{\mu_1}{\mu_2}} x_{\mu_3} \ldots x_{\mu_n} 
\, + \, ( \half n(n-1) - 1 ) ~ \mbox{terms} ] ~+~ O(x^{n-4}) \right] 
\label{gdeco} 
\end{equation}  
In this and similar decompositions we will omit the overall power of the 
factor of $(x^2)$ as it plays no significant role in the derivation and can 
readily be restored by simple dimensional analysis. Here $A(\alpha,n)$ and 
$B(\alpha,n)$ are invariant amplitues and our aim is to determine a recurrence 
relation involving $A(\alpha,n)$ only. This is because after contracting 
(\ref{gdeco}) with $\Delta^{\mu_1} \ldots \Delta^{\mu_n}$ only $A(\alpha,n)$ 
remains which is the value of $G_{00n}(\alpha)$. The amplitude $B(\alpha,n)$ is
necessary to achieve this as can be seen in the explicit derivation, though the
terms in powers of $x$ will not be necessary and these have been indicated by 
the order symbol. In the following construction of a relation for $A(\alpha,n)$
we consider only the leading contribution in powers of $x$ when we take two  
contractions. These are $\eta^{\mu_{n-1}\mu_n}$ and $x^{\mu_n}$ and give 
respectively  
\begin{eqnarray} 
A(\alpha-1,n-2) &=& A(\alpha,n) ~+~ 2(\mu+n-2) B(\alpha,n) \nonumber \\ 
2 G^{0111}_{00(n-1)} &=& A(\alpha,n) ~+~ (n-1)B(\alpha,n) 
\end{eqnarray} 
where $G^{0111}_{pqn}(\alpha)$ corresponds to the integral of fig. 3 but 
with the exponent of the top left line replaced by zero. In other words it
is an elementary chain integral of the form which has arisen already. From
these two expressions we deduce, for arbitrary $\alpha$,  
\begin{equation} 
G_{00(n-2)}(\alpha-1) ~=~ \frac{4(\mu+n-2)}{(n-1)} G^{0111}_{00(n-1)}(\alpha) 
{}~-~ \frac{(2\mu-3+n)}{(n-1)}G_{00n}(\alpha) 
\label{rr1} 
\end{equation} 
from the decomposition the various integrals we require are 
$G_{00n}(2\mu-p+\Delta)$ for $p$ $=$ $2$, $3$ and $4$. Indeed once we have 
obtained one of these values the others will follow from elementary application
of (\ref{rr1}). It turns out that $G_{00n}(2\mu-3+\Delta)$ is $\Delta$-finite.
The justification for this will prove to be fairly general and will be used
later in solving recurrence relations for other integrals. Essentially it
follows from the fact that the dimension of the integral is greater than unity.
We first recall the usual calculational technique used in \cite{12} for 
computing integrals without a moment factor in their numerators but which are 
divergent and similar in form to fig. 3. The first step is to transform it to 
the momentum representation defined in \cite{12}. In this one has to include 
factors arising from the Fourier transform. In particular one factor is related
to the dimension of the integral. If this has the form $-$ $(p$ $+$ $\Delta)$ 
for a positive integer $p$, then the corresponding factor is divergent with 
respect to $\Delta$. It therefore remains only to compute the momentum 
representation integral for $\Delta$ $=$ $0$. This is possible as the final 
form is related, for example by integration by parts, to known integrals like 
those of Chetyrkin and Tkachov, \cite{24}, denoted by $ChT(\alpha,\beta)$, 
\cite{12}. It follows therefore for this case that if the original integral has
a dimension greater than unity then it is finite. For the integrals with a 
non-zero moment the argument is similar. Indeed we checked this reasoning by 
explicitly computing $G_{00n}(2\mu-3+\Delta)$ for various values of low $n$. 
For future reference we make some comment on the extension of this argument to 
other integrals not necessarily of the exponent structure of fig. 1. It may not
always be the case that the corresponding $n$ $=$ $0$ integral is 
$\Delta$-finite even though its dimension satisfies the above criterion. 
However, its moment partner may be. This is because the presence of a numerator
factor like $\Delta(y-z)$ alters the nature of the singularity which we are 
interested in. These remarks can be simply illustrated by considering the 
elementary function $a_n(\alpha)$ which is common in all our calculations. For 
example, the function $a_0(\mu-\Delta)$ is divergent in the limit $\Delta$ 
$\rightarrow$ $0$, whereas for a non-zero $n$ $a_n(\mu-\Delta)$ is finite in 
the same limit. The presence of the non-zero moment appears to be regularizing 
the infinity of the $n$ $=$ $0$ function and it is this which occurs in the two
and three loop integrals. Indeed throughout our application of recurrence 
relations we always have used this dimensional argument for $n$ $\neq$ $0$ 
integrals to determine the terminating case. For all the cases we are 
interested in we checked the finiteness explicitly for at least one value of 
non-zero $n$. The argument appears robust. 

Therefore given this property we can deduce $G_{00n}(2\mu-4+\Delta)$, for 
example, from (\ref{rr1}) which now is reduced to a computation of chain 
integrals. Again we have checked this result by carrying out the explicit
calculation for low $n$ and found agreement. To complete the calculation of
the second graph a recurrence relation for $G_{20n}(\alpha)$ is deduced in 
the same way as for (\ref{rr1}). The major difference here is that the 
decomposition analogous to (\ref{gdeco}) now has four terms to $O(x^{(n-2)})$.
However there are various contractions which can be made from which one can
deduce, for example, $G_{20n}(\alpha)$. Finally the two projections of  
(\ref{proj}) can be performed and the contributing value of the second integral
to $\eta_{{\cal O}_g}$ is  
\begin{eqnarray} 
&-& [8\mu^4 n^2 + 8\mu^4 n + 8\mu^4 + \mu^3 n^4 + 2\mu^3 n^3 - 11\mu^3 n^2 
\nonumber \\ 
&&~ - 12\mu^3 n - 24\mu^3 - 3\mu^2 n^4 - 6\mu^2 n^3 - 19\mu^2 n^2 - 16\mu^2 n
\nonumber \\ 
&&~ + 24\mu^2 - \mu n^6 - 3\mu n^5 - 5\mu n^4 - 5\mu n^3 + 34\mu n^2 + 36\mu n 
\nonumber \\ 
&&~ - 8\mu + 2n^6 + 6n^5 + 10n^4 + 10n^3 - 12n^2 - 16n] \nonumber \\
&&~~ \times \mu \Gamma(\mu-1) \Gamma(n+1-\mu) [C_2(R) - C_2(G)/2] 
\eta_1^{\mbox{o}} \nonumber \\ 
&&~~~ /[n(n^2-1)(n+2) \Gamma(3-\mu) \Gamma(\mu+n+1) T(R)] 
\end{eqnarray} 
We close this section with the remark that we have made use of the computer 
algebra and symbolic manipulation packages {\sc Reduce}, \cite{25}, and {\sc 
Form}, \cite{26}, in this and subsequent graph evaluations as there is a 
substantial amount of tedious algebra to handle. 

\sect{$C_2(G)$ sector.} 

We now detail the calculation of the remaining two graphs of fig. 2 each of
which involves only the colour group factor $C_2(G)$. The strategy is the same 
as in the $C_2(R)$ sector. In other words we take the trace over the fermion 
loop and evaluate the pole part of the resulting bosonic integrals. The major 
effort in the calculation resides in the determination of the integrals which 
are not reducible to chains. First, we consider the two loop graph. The three 
gluon insertion Feynman rules can be found in, for example, \cite{2}. It has a 
part which involves a sum over the operator moment and it is therefore not 
clear whether it will be possible to obtain a closed analytic expression for 
the contribution from this graph. Moreover, the distribution of the exponents
around the two loop topology is different from that of fig. 3 and so a new
recurrence relation will be needed. The integral is defined in fig. 4 where
$\alpha$ and $\beta$ are arbitrary exponents and $n$ and $p$ are integers. In
addition to the simple chain diagrams given by the graph of fig. 4 where each
of the lines with a unit exponent is successively replaced by zero, we have to 
deal with those graphs where the exponents of each of the lines are also 
replaced by an exponent of $(-$ $1)$. These integrals are easily related to a
sum of chain diagrams by integration rules like that illustrated in fig. 5
which is simple to establish. As these resulting chain integrals are elementary
to deal with we focus on the construction of a relation for 
$H_{pn}(\alpha,\beta)$ and consider $H_{0n}(\alpha,\beta)$ first. A clue on how
to proceed comes from the case $H_{00}(\alpha,\beta)$ which has been determined
in closed form in \cite{24}. It is related to $ChT(\mu-\alpha,\mu-\beta)$ of 
\cite{12} after applying the momentum representation transformation of 
\cite{12}. The presence of a non-zero moment $n$ however, disrupts the 
construction of a closed form when the same algorithm is applied. Instead we 
apply a series of transformations to $H_{0n}(\alpha,\beta)$. Noting that under 
a conformal transformation $z^\mu$ $\rightarrow$ $z^\mu/z^2$ and therefore 
$\Delta z$ $\rightarrow$ $\Delta z/z^2$, we apply a left conformal 
transformation on fig. 4 with the origin at the left external point. The lines 
joining the upper vertex now have no moment dependence and so we can apply a 
$\nearrow$ transformation of \cite{12}. Finally a momentum representation 
transformation leaves us with an integral proportional to the graph given in 
fig. 6. Again as the lower vertex has no lines joining it which have a moment 
dependence we can integrate by parts on it using, for example, the rule of 
\cite{23} and relate it to four integrals. Three of these are simple chains 
whilst the fourth is related to $H_{0n}(\alpha,\beta-1)$ when one undoes the 
original  transformations. The result therefore is, for arbitrary $\alpha$ and 
$\beta$, 
\begin{eqnarray} 
H_{0n}(\alpha,\beta) &=& \frac{(2\mu-\alpha-\beta+n-1)(\alpha+\beta-\mu)}
{(\beta-1)(\mu-\beta-1)} H_{0n}(\alpha,\beta-1) \nonumber \\ 
&-& \frac{a(1)a(\beta+2-\mu)a_n(3\mu-\alpha-\beta-2+n)}{(\beta-1)a(3-\mu) 
a_n(2\mu-\alpha-2+n)} \nonumber \\ 
&& ~ \times \left[ \frac{(\beta-1)}{(\mu-\beta-1)} \nu(2,\beta,2\mu-\beta-2) 
{}~+~ \nu_{non}(\alpha,2,2\mu-\alpha-2+n) \right] 
\label{rr2} 
\end{eqnarray}  
For (\ref{rr2}) to be practical we once again need to have a terminating value
which will be an integral which is finite with respect to the regularization.
Unlike in section 3 we require several distinct series of expressions such as
$H_{0n}(\mu-1+\Delta,\mu-2+\Delta)$, $H_{0n}(2\mu-3+\Delta,\Delta)$ and  
$H_{0n}(2\mu-3+\Delta,\mu-1+\Delta)$. For the former the natural endpoint is  
$H_{0n}(\mu-1+\Delta,\mu-1+\Delta)$ $=$ $O(1)$. This has been verified for
various values of small $n$ and simple substitution reveals  
\begin{equation} 
H_{0n}(\mu-1+\Delta,\mu-2+\Delta) ~=~ - ~ \frac{(2\mu-3)(\mu-1)} 
{2(n+1)(\mu-1+n)} \frac{\nu(1,1,2\mu-2)}{\Gamma(\mu)\Delta} 
\label{res12} 
\end{equation} 
which we have also checked for low $n$. Of course we have only recorded the 
value of the $O(1/\Delta)$ part of the integral since the contribution to 
$\eta_{{\cal O}_g}$ only arises from the residue of the simple pole. One cannot
always directly use (\ref{rr2}) for each integral due to the fact that if, for 
instance, $\beta$ $=$ $\mu$ $-$ $1$ $+$ $\Delta$ then a denominator factor of 
(\ref{rr2}) is $\Delta$. This would require the associated two loop integral to
be computed to the next order in $\Delta$ which would be tedious. Instead one 
can employ the identity 
\begin{equation} 
H_{0n}(\mu-2+\Delta,\mu-1+\Delta) ~=~ \sum_{r=0}^n 
\left( 
\begin{array}{c} 
n \\ r \\  
\end{array} 
\right) 
(-1)^r H_{0r}(\mu-1+\Delta,\mu-2+\Delta) 
\end{equation} 
and insert the previous result. Elementary algebraic identities allow the 
finite sum to be determined, leaving 
\begin{equation} 
H_{0n}(\mu-2+\Delta,\mu-1+\Delta) ~=~ - \, 
\frac{(2\mu-3)(\mu-1)\nu(1,1,2\mu-2)}{2(\mu-2)\Gamma(\mu)\Delta} 
\left[ \frac{1}{(n+1)} ~-~ \frac{\Gamma(\mu-1)\Gamma(n+1)}{\Gamma(\mu+n)} 
\right] 
\end{equation} 
For the other structures we note that by our earlier dimensional argument 
$H_{0n}(2\mu-3+\Delta,1+\Delta)$ and $H_{0n}(2\mu-2+\Delta,\Delta)$ are 
$\Delta$-finite and lead to the following results  
\begin{eqnarray} 
H_{0n}(2\mu-3+\Delta,\Delta) &=&  
\frac{\nu_{00n}(1,2,2\mu-3)}{2(n+1)\Gamma(\mu+n)\Delta} \nonumber \\  
H_{0n}(2\mu-4+\Delta,1+\Delta) &=&  
\frac{(2\mu-3)\nu(1,1,2\mu-2)}{2(\mu-2)(n+1)\Gamma(\mu+n)\Delta} 
\left[ \frac{\Gamma(4-\mu+n)}{\Gamma(3-\mu)} ~-~ \Gamma(n+2) \right] 
\nonumber \\ 
H_{0n}(2\mu-2+\Delta,-1+\Delta) &=&  
- \, \frac{(\mu-1)(2\mu-3)\nu_{00n}(1,2,2\mu-2)}{2(n+1)\Gamma(\mu+n)\Delta} 
\end{eqnarray} 
Again we have verified these (and others) all explicitly for at least one 
non-zero value of $n$. It turns out that the remaining integrals which we need
of the form $H_{0n}(2\mu-p+\Delta,\mu-q+\Delta)$ for positive integers $p$ and
$q$ are $\Delta$-finite.  

To complete the construction of results for the topology of fig. 4 we note that
applying the generalized moment integration by parts rule of \cite{23} to 
$H_{(p-1)(n+1)}(\alpha,\beta)$ gives the reduction formula, for arbitrary 
$\alpha$ and $\beta$,  
\begin{eqnarray} 
H_{pn}(\alpha,\beta) &=& \frac{(2\mu-\alpha-\beta+n-1)}{(n+1)} 
H_{(p-1)(n+1)}(\alpha,\beta) \nonumber \\ 
&& -~ \frac{\alpha}{(n+1)} \left[ H_{(p-1)(n+1)}^{101}(\alpha+1,\beta) 
- H_{(p-1)(n+1)}^{011}(\alpha+1,\beta) \right] \nonumber \\  
&& -~ \frac{\beta}{(n+1)} \left[ H_{(p-1)(n+1)}^{101}(\alpha,\beta+1) 
- H_{(p-1)(n+1)}^{110}(\alpha,\beta+1) \right] 
\label{rr3} 
\end{eqnarray} 
where the superscripts on an $H$-function here refer to the exponents 
respectively of the top left, central and top right lines relative to the 
diagram of fig. 4. The result (\ref{rr3}) allows us to reduce any integral
$H_{pn}(\alpha,\beta)$ with $p$ $>$ $0$ down to ones of the form 
$H_{0n}(\alpha,\beta)$ whose values are known from (\ref{rr2}). Therefore one 
can now assemble all the pieces to both projections of the original two loop 
diagram. 

For that part of the operator insertion which does not involve a finite sum 
over moments one obtains the contribution 
\begin{eqnarray} 
&& \left[ (4\mu^3n^2 - 4\mu^3 - 10\mu^2n^2 + 2\mu^2n + 8\mu^2 + 8\mu n^2 
- 4\mu n - 3\mu - 2n^2 + 2n)\Gamma(\mu + n) \right. \nonumber \\ 
&& \left. ~~ -~ (2\mu^2n - 2\mu^2 - 3\mu n + 4\mu + n - 1) \Gamma(\mu + 1) 
\Gamma(n+1) \right] \eta^{\mbox{o}}_1 C_2(G) \nonumber \\ 
&& ~~~~ /[2(2\mu - 1)(\mu - 1)(\mu - 2)(n - 1)\Gamma(\mu + n)n]  
\end{eqnarray} 
However, the total of all pieces to that part which does involve the sum 
yields, ignoring colour factors and powers of $z$ for the moment,  
\begin{eqnarray} 
&-& \frac{4(\mu-1)\nu(1,1,2\mu-2)}{(2\mu-1)\Gamma(\mu)} \sum_{r=1}^{n-2} 
\frac{1}{r(r+1)\Gamma(\mu+r+1)} \nonumber \\ 
&& \times \left[ [ 4r(\mu-1)^2 + 1] \Gamma(\mu+r+1) ~-~ 
[ r(2\mu-1)(\mu-1) + \mu)]\Gamma(\mu)\Gamma(r+2) ] \right] 
\label{cgsum} 
\end{eqnarray}  
From the definition for $S_l(n)$ and the result 
\begin{equation} 
\sum_{r=1}^n \frac{\Gamma(r+a)}{\Gamma(r+b)} ~=~ 
\frac{1}{(a-b-1)} \left[ \frac{\Gamma(a+1)}{\Gamma(b)} ~-~ 
\frac{\Gamma(n+a+1)}{\Gamma(n+b)} \right] 
\end{equation} 
then we deduce the closed form for (\ref{cgsum}) is  
\begin{eqnarray} 
&-& \frac{4(\mu-1)\nu(1,1,2\mu-2)}{(2\mu-1)\Gamma(\mu)} \left[ 
(2\mu-3)(2\mu-1) \left( S_1(n) - \frac{(n+1)}{n} \right) \right. \nonumber \\ 
&& \left. ~~ + S_1(n) - \frac{(2n-1)}{n(n-1)} - 2 
+ [ (2\mu-1)(n-1) + 1 ] \frac{\Gamma(n-1)\Gamma(\mu)}{\Gamma(\mu+n-1)} \right] 
\end{eqnarray}  
Thus the total contribution from this two loop graph is 
\begin{equation} 
\left[ \frac{2\mu(\mu - 1)S_1(n)}{(2\mu - 1)(\mu - 2)}  
{} ~+~ \frac{(2\mu n - n + 1)\Gamma(\mu+1)\Gamma(n)} 
{2(2\mu - 1)(\mu - 1)(\mu - 2)\Gamma(\mu + n)}   
{} ~-~ \frac{(2\mu^2n - 3\mu n + \mu + 2n)} 
{2(2\mu - 1)(\mu - 1)(\mu - 2)n} \right] \frac{\eta_1^{\mbox{o}} C_2(G)}{T(R)}  
\end{equation} 

To complete our computation we now turn to the final graph of fig. 2, which is 
the hardest of the four to evaluate. As we are in effect performing two 
separate projections on the integral itself, we note that multiplying the graph
by $\Delta^\mu p^\nu$, for instance, allows us to immediately complete the 
integration for the quark loop that the $p^\nu$ contracts into. The resulting 
two loop integral is effectively completed since its constituent scalar 
integrals have already been evaluated for the third graph of fig. 2. A similar 
property arises for the $\eta^{\mu\nu}$ contraction of the second term of the 
operator insertion (\ref{oprule}). As for the $\Delta^\mu p^\nu$ contraction 
there is also no new feature for this case. Instead we consider the remaining 
two terms of (\ref{oprule}) in fig. 2 and their $\eta^{\mu\nu}$ projection. In 
taking the relevant traces one obtains scalar integrals which are now either 
simple chains or two loop integrals which have already been considered. In 
addition we obtain a new three loop topology whose most general form is given 
in fig. 7 and has a similar integral definition to (\ref{gdefn}). Here $p$, 
$q$, $n$ and $\gamma$ are integers with $\alpha$ and $\beta$ arbitrary 
exponents. We emphasise that here the line with exponent $( \, -$ $\gamma)$ 
corresponds to a numerator factor of $[(y-z)^2]^\gamma$. Similar to the other 
$C_2(G)$ two loop chain type graphs there are a variety of three loop graphs of
the form of fig. 7 but where one of the lines with unit exponent is replaced by
$( \, -$ $1)$ for $\gamma$ $=$ $1$, $2$ or $3$. These can readily be reduced to
at most two loop graphs already computed using integration rules similar to 
that of fig. 5. 

Therefore since the structure of fig. 7 is new to the calculation we give some 
details of its treatment, though the algorithm is no different from those 
already established for the two loop cases. We consider the construction of a 
relation to deduce $C_{00\gamma}(\alpha,\beta,n)$ first and note that its 
decomposition into Lorentz invariants where the $\Delta^\mu$ vectors have been 
omitted is the same as in (\ref{gdeco}). However, due to the form of the 
present integral the contractions $\eta^{\mu_{n-1}\mu_n}$ and $x^{\mu_n}$ yield 
respectively  
\begin{equation} 
A(\alpha-1,\beta,n-2) ~=~ A(\alpha,\beta,n) ~+~ 2(\mu+n-2)B(\alpha,\beta,n)  
\end{equation} 
\begin{equation} 
A(\alpha,\beta,n) ~+~ (n-1)B(\alpha,\beta,n) ~=~ 
\half [ A(\alpha,\beta,n-1) ~+~ A(\alpha-1,\beta,n-1) ~-~ 
A(\alpha,\beta-1,n-1) ] \nonumber  
\end{equation} 
Eliminating $B(\alpha,\beta,n)$ gives 
\begin{eqnarray} 
A(\alpha-1,\beta,n-2) &=&  -~ \frac{(2\mu+n-3)}{(n-1)} A(\alpha,\beta,n)  
\nonumber \\ 
&+& \frac{(\mu+n-2)}{(n-1)} 
[ A(\alpha,\beta,n-1) ~+~ A(\alpha-1,\beta,n-1) ~-~ A(\alpha,\beta-1,n-1) ] 
\nonumber \\ 
\label{intrr} 
\end{eqnarray} 
This is not quite in a form which is practical since we require a relation
between $A(\alpha,\beta,n)$ and say $A(\alpha-1,\beta,n)$ where the former set
can be set to zero upon reaching a terminating value. The presence of the final
two terms in the square brackets leaves us an equation which will relate 
several integrals without allowing us to evaluate them individually. To 
eliminate these problem terms we consider the general integral of fig. 7 and
apply a conformal left transformation to it in the language of \cite{12}. Then 
integrating by parts on the central $4$-vertex with the line with exponent 
$\beta$ as reference and applying the inverse conformal left transformation to 
the resulting five integrals, we obtain the general result, for arbitrary 
$\alpha$ and $\beta$,   
\begin{eqnarray} 
C_{pq\gamma}(\alpha,\beta,n) &=& \frac{(2\mu-\alpha-\beta-2+n)}{(\alpha-\beta)}
[ C_{pq\gamma}(\alpha,\beta-1,n) - C_{pq\gamma}(\alpha-1,\beta,n) ] 
\nonumber \\ 
&& +~ \frac{1}{(\alpha-\beta)} 
[ C^{021}_{pq\gamma}(\alpha,\beta-1,n) - C^{120}_{pq\gamma}(\alpha-1,\beta,n) ] 
\nonumber \\ 
&& +~ \frac{1}{(\alpha-\beta)} 
[ C^{021}_{qp\gamma}(\alpha,\beta-1,n) - C^{120}_{qp\gamma}(\alpha-1,\beta,n) ] 
\nonumber \\ 
&& +~ \frac{n}{(\alpha-\beta)} C_{pq\gamma}(\alpha-1,\beta,n-1) 
\label{confrr} 
\end{eqnarray} 
where a similar notation to (\ref{rr3}) is used but here the superscripts refer
only to the exponents of the lines joining the top vertex. Although we are 
dealing with the leading term $A(\alpha,\beta,n)$ in (\ref{intrr}) we can use 
(\ref{confrr}) to eliminate the unnecessary terms since after the contraction 
with $\Delta^{\mu_1} \ldots \Delta^{\mu_n}$ only $A(\alpha,\beta,n)$ will 
survive and be equal in value to $C_{00\gamma}(\alpha,\beta,n)$. Consequently 
after some elementary algebra which includes shifting exponents, we deduce, for
arbitrary $\alpha$ and $\beta$,   
\begin{eqnarray} 
C_{00\gamma}(\alpha,\beta,n) &=& -~ \frac{(2\mu+n-1)(2\mu-\alpha-\beta+n-2)}
{(n+1)(\mu-\alpha-\beta-2)} C_{00\gamma}(\alpha+1,\beta,n+2) \nonumber \\  
&& +~ \frac{(\mu+n)(2\mu-2\alpha+n-3)}{(n+1)(\mu-\alpha-\beta-2)} 
C_{00\gamma}(\alpha+1,\beta,n+1) \nonumber \\  
&& +~ \frac{2(\mu+n)}{(n+1)(\mu-\alpha-\beta-2)} 
[ C^{021}_{00\gamma}(\alpha+1,\beta-1,n+1) 
- C^{120}_{00\gamma}(\alpha,\beta,n+1) ] \nonumber \\  
\end{eqnarray}  
This differs in a minor respect from the two loop relation of (\ref{rr2}) in
that the first two terms on the right side are not at the same moment level. 
However, the dimension argument we use to find a terminating value is valid 
for all $n$ $\geq$ $1$ and these terms will be $\Delta$-finite.  

Relations are also required for $C_{10\gamma}(\alpha,\beta,n)$, 
$C_{11\gamma}(\alpha,\beta,n)$ and $C_{20\gamma}(\alpha,\beta,n)$ and these
are determined by the same method as for $C_{00\gamma}(\alpha,\beta,n)$ though
there are of course more invariant amplitudes which can appear. However, one
constructs the result for $C_{10\gamma}(\alpha,\beta,n)$ first since the 
decomposition of the other two relies on this lower level expression and 
$C_{00\gamma}(\alpha,\beta,n)$. For completeness we quote the recurrence 
relations for each level in appendix A where again the integrals where a line 
is zero are easily computed. As an aid to an interested reader we record the 
explicit results of the recurrence relations for a set of representative 
integrals at each level, noting that we have used the obvious values such as 
$C_{000}(2\mu-3+\Delta,\mu-1+\Delta,n)$ $=$ $O(1)$ for all $n$ $\geq$ $1$ to 
terminate the relations. We found   
\begin{eqnarray} 
C_{003}(2\mu-2+\Delta,\mu-1+\Delta,n)  
&=& - \, \left[ \frac{}{} [2(2\mu - 3)(\mu^2 - 1) \Gamma(\mu + n + 1) 
\right. \nonumber \\ 
&& \left. ~~~~~~ - \Gamma(\mu+2)\Gamma(n+3) ] a(2\mu - 2) \right. 
\nonumber \\ 
&& \left. ~~~ + 2(2\mu - 3)(\mu - 1) a_n(2\mu - 2)\Gamma(\mu+2) 
\frac{}{} \right] \nonumber \\ 
&& ~~~~ \times a(2\mu - 2)a^4(1)\mu/[4(2\mu - 1)^2(\mu - 2)] \nonumber \\ 
&& ~~~~~ /[(n + 2)(n + 1) \Gamma(\mu + n)\Gamma(\mu)\Delta] 
\end{eqnarray} 
\begin{eqnarray} 
C_{100}(2\mu-4+\Delta,\mu-2+\Delta,n)  
&=& \left[ \frac{}{} (\mu^2 + 2\mu n^2 + 8\mu n + 4\mu - 3n^2 - 12n - 8)
(\mu - n - 2) \right. \nonumber \\ 
&& \left. ~~ \times (\mu - n - 3)a_n(2\mu - 2)\Gamma(\mu) \right. \nonumber \\ 
&& \left. ~+~ (\mu - 1)^2(\mu - 2)(n + 2)a(2\mu - 2)\Gamma(\mu + n + 1) 
\frac{}{} \right] \nonumber \\ 
&& ~ \times (2\mu - 3)a(2\mu - 2) \Gamma^3(\mu)/[2(\mu - 1)^4(\mu - 2)^2] 
\nonumber \\ 
&& ~~ /[(n + 3)(n + 2)(n + 1)\Gamma(\mu + n + 1)\Delta] 
\end{eqnarray} 
\begin{eqnarray} 
C_{112}(2\mu-3+\Delta,\mu-1+\Delta,n) 
&=& \left[ \frac{}{} [ (2\mu^3n^2 + 10\mu^3n + 10\mu^3 - 3\mu^2n^2 - 19\mu^2n 
- 25\mu^2 \right. \nonumber \\ 
&& \left. ~~~~ - \mu n^2 - \mu n + 7\mu + n^2 + 7n + 6) \Gamma(\mu + n + 3) 
\right. \nonumber \\ 
&& \left. ~ - (\mu^2n + 2\mu n^2 + 4\mu n + \mu - n^2 - 3n - 2)(\mu - 1) 
\right. \nonumber \\ 
&& \left. ~~~~~ \times \Gamma(\mu) \Gamma(n+5) ] 
a(2\mu - 2) \right. \nonumber \\ 
&& \left. ~ + (\mu^4 + 2\mu^3n^2 + 8\mu^3n + 4\mu^3 + 4\mu^2n^4 + 36\mu^2n^3 
\right. \nonumber \\ 
&& \left. ~~~~~ + 110\mu^2n^2 + 132\mu^2n + 53\mu^2 - 4\mu n^4 - 38\mu n^3 
\right. \nonumber \\ 
&& \left. ~~~~~ - 124\mu n^2 - 162\mu n - 70\mu + n^4 + 10n^3 + 35n^2 
\right. \nonumber \\ 
&& \left. ~~~~~ + 50n + 24)(2\mu - 3) (\mu - n - 2)a_n(2\mu - 2)\Gamma(\mu) 
\frac{}{} \right] \nonumber \\ 
&& ~ \times a(2\mu - 2)\Gamma^3(\mu)\mu /[4(2\mu - 1)^2(\mu - 1)^4(\mu - 2)] 
\nonumber \\ 
&& ~~ /[(n + 4)(n + 3)(n + 2)(n + 1)\Gamma(\mu + n + 2)\Delta] 
\end{eqnarray} 
and 
\begin{eqnarray} 
C_{201}(2\mu-4+\Delta,\mu-1+\Delta,n) 
&=& \left[ \frac{}{} [ (\mu^3 + 4\mu^2n + 5\mu^2 + 5\mu n^2 + 13\mu n + 8\mu 
- 2n^2 \right. \nonumber \\  
&& \left. ~~~ - 6n - 4)(n + 4)(n + 3)\Gamma(\mu)\Gamma(n+2) \right. \nonumber \\
&& \left. ~~~ - 2(\mu n + 3\mu - 1) \Gamma(\mu + n + 3)\mu ] (n + 2)a(2\mu - 2) 
\right. \nonumber \\ 
&& \left. ~ - 4(\mu^3 + 2\mu^2n^2 + 10\mu^2n + 9\mu^2 + 2\mu n^4 + 20\mu n^3 
\right. \nonumber \\ 
&& \left. ~~~~~ + 69\mu n^2 + 95\mu n + 44\mu - n^4 - 10n^3 - 35n^2 - 50n 
\right. \nonumber \\ 
&& \left. ~~~~~ - 24) (\mu - n - 2)(\mu - n - 3)a_n(2\mu - 2) 
\Gamma(\mu) \frac{}{} \right] \nonumber \\ 
&& ~ \times a(2\mu - 2)\Gamma^3(\mu-1)/[4(2\mu - 1)(\mu - 2)(n + 4)] 
\nonumber \\ 
&& ~~ /[(n + 3)(n + 2)(n + 1) \Gamma(\mu + n + 2)\Delta] 
\label{c20141n} 
\end{eqnarray} 

With the results from these recurrence relations and the earlier two loop 
results we summarise this intermediate step in determining the value of the
three loop graph by giving the contributions from the $\eta^{\mu\nu}$ 
projections of the first and third terms of the operator insertion, 
(\ref{oprule}). These are respectively  
\begin{eqnarray}  
&-& [16\mu^6 - 16\mu^5n - 104\mu^5 - 4\mu^4n^2 + 60\mu^4n + 264\mu^4 
+ 16\mu^3n^2 - 80\mu^3n - 336\mu^3 - 21\mu^2n^2 \nonumber \\ 
&& + \, 51\mu^2n + 232\mu^2 + 11\mu n^2 - 20\mu n - 88\mu - 2n^2 + 4n + 16] 
\Gamma(n+2-\mu) \Gamma(\mu - 1) \nonumber \\ 
&& ~ \times \mu \eta_1^{\mbox{o}} C_2(G)/[8(2\mu - 1)(\mu - 1)^2(\mu - 2)^2 
(n + 2)(n + 1) \Gamma(2-\mu))\Gamma(\mu + n) T(R)] \nonumber \\ 
&-& (4\mu - 1)\Gamma(\mu - 1)\Gamma(n+1)\mu \eta_1^{\mbox{o}} C_2(G) 
/[4(2\mu - 1)(\mu - 2) \Gamma(\mu + n) T(R)] \nonumber \\ 
&-& [16\mu^6 + 16\mu^5n - 88\mu^5 - 4\mu^4n^2 - 84\mu^4n + 192\mu^4 
+ 12\mu^3n^2 + 164\mu^3n - 216\mu^3 - 13\mu^2n^2 \nonumber \\ 
&& - \, 149\mu^2n + 136\mu^2 + 8\mu n^2 + 63\mu n - 48\mu - 4n^2 - 10n + 8] \mu
\eta_1^{\mbox{o}} C_2(G) \nonumber \\  
&& ~/[8(2\mu - 1)(\mu + n - 1)(\mu - 1)^3(\mu - 2)(n + 2)(n + 1) T(R)] 
\label{glu5} 
\end{eqnarray}  
and 
\begin{eqnarray}  
&+& [16\mu^7n^2 + 48\mu^7n + 32\mu^7 + 32\mu^6n^3 - 56\mu^6n^2 - 360\mu^6n 
- 256\mu^6 + 16\mu^5n^4 - 160\mu^5n^3 \nonumber \\ 
&& - \, 72\mu^5n^2 + 1048\mu^5n + 852\mu^5 - 88\mu^4n^4 + 312\mu^4n^3 
+ 614\mu^4n^2 - 1518\mu^4n - 1530\mu^4 \nonumber \\ 
&& + \, 184\mu^3n^4 - 288\mu^3n^3 - 1145\mu^3n^2 + 1141\mu^3n + 1600\mu^3 
- 182\mu^2n^4 + 108\mu^2n^3 \nonumber \\ 
&& + \, 973\mu^2n^2 - 401\mu^2n  - 970\mu^2 + 83\mu n^4 + 4\mu n^3 - 385\mu n^2
+ 34\mu n + 312\mu - 14n^4 \nonumber \\ 
&& - \, 8n^3 + 54n^2 + 8n - 40]\Gamma(n+1-\mu)\Gamma(\mu - 1)\mu 
\eta_1^{\mbox{o}} C_2(G) \nonumber \\ 
&& ~ /[8(2\mu - 1)(\mu - 1)^2(\mu - 2)^2(n + 2)(n^2 - 1)\Gamma(2-\mu) 
\Gamma(\mu + n)n T(R)] \nonumber \\ 
&-& (2\mu^2 + 3\mu n - 3\mu - n + 1)\Gamma(\mu - 1)\Gamma(n-1)\mu 
\eta_1^{\mbox{o}} C_2(G) /[2(2\mu - 1)(\mu - 2)\Gamma(\mu + n) T(R)] 
\nonumber \\ 
&-& [16\mu^6n^2 + 48\mu^6n + 32\mu^6 + 16\mu^5n^3 - 40\mu^5n^2 - 232\mu^5n 
- 192\mu^5 + 4\mu^4n^4 - 64\mu^4n^3 \nonumber \\ 
&& - \, 48\mu^4 n^2 + 388\mu^4n + 460\mu^4 - 12\mu^3n^4 + 68\mu^3n^3 
+ 200\mu^3n^2 - 236\mu^3n - 558\mu^3 \nonumber \\ 
&& + \, 5\mu^2n^4 - 4\mu^2n^3 - 166\mu^2n^2 - 15\mu^2n + 356\mu^2 + 8\mu n^4 
- 16\mu n^3 + 33\mu n^2 + 61\mu n \nonumber \\ 
&& - \, 110\mu - 4n^4 + 6n^2 - 14n + 12] \mu \eta_1^{\mbox{o}} C_2(G) 
\nonumber \\ 
&& ~ /[8(2\mu - 1)(\mu + n - 1)(\mu - 1)^3(\mu - 2)(n + 2)(n^2 - 1)n T(R)] 
\label{glu6} 
\end{eqnarray}  
Therefore summing all the contributions from each of the projections the full 
contribution from the final graph of fig. 2 is simply  
\begin{eqnarray}  
&+& [32\mu^7n^2 + 32\mu^7n + 32\mu^7 - 96\mu^6n^2 - 96\mu^6n - 160\mu^6 
+ 28\mu^5n^4 + 56\mu^5n^3 - 20\mu^5n^2 \nonumber \\  
&& - \, 48\mu^5 n + 324\mu^5 - 148\mu^4n^4 - 296\mu^4n^3 + 394\mu^4n^2 
+ 542\mu^4n - 338\mu^4 + 4\mu^3n^6 \nonumber \\ 
&& + \, 12\mu^3n^5 + 303\mu^3n^4 + 586\mu^3n^3 - 606\mu^3n^2 - 897\mu^3n 
+ 188\mu^3 -  8\mu^2n^6 - 24\mu^2n^5 \nonumber \\ 
&& - \, 288\mu^2n^4 - 536\mu^2n^3 + 421\mu^2n^2 
+ 685\mu^2n - 50\mu^2 + 5\mu n^6 + 15\mu n^5 + 125\mu n^4 \nonumber \\ 
&& + \, 225\mu n^3 - 144\mu n^2 - 254\mu n + 4\mu - n^6 - 3n^5 - 19n^4 - 33n^3 
+ 20n^2 + 36n ] \nonumber \\ 
&& ~ \times \Gamma(n+1-\mu)\Gamma(\mu - 1)\mu \eta_1^{\mbox{o}} C_2(G) 
\nonumber \\ 
&& ~ /[8(2\mu - 1)(\mu - 1)^2(\mu - 2)(n + 2) 
(n^2 - 1)\Gamma(2-\mu)\Gamma(\mu + n + 1)n T(R)] \nonumber \\ 
&-& (2\mu n - n + 1)\Gamma(\mu - 1)\Gamma(n)\mu \eta_1^{\mbox{o}} C_2(G) 
/[2(2\mu - 1)(\mu - 2)\Gamma(\mu + n) T(R)] \nonumber \\ 
&-& [32\mu^5n^2 + 32\mu^5n + 32\mu^5 - 144\mu^4n^2 - 144\mu^4n - 160\mu^4 
+ 236\mu^3n^2 + 236\mu^3n + 316\mu^3 \nonumber \\ 
&& - \, 4\mu^2n^3 - 172\mu^2n^2 - 160\mu^2n - 298\mu^2 + 8\mu n^3 
+ 55\mu n^2 + 31\mu n + 130\mu - 4n^3 - 6n^2 \nonumber \\ 
&& + \, 6n - 20 ] \mu \eta_1^{\mbox{o}} C_2(G)/[8(2\mu - 1)(\mu - 1)^3(\mu - 2) 
(n + 2)(n^2 - 1)n T(R)] 
\end{eqnarray}  

\sect{Results and discussion.} 

We are now in a position to complete the calculation by adding the 
contributions from the four graphs of fig. 2 as well as including the gluon
anomalous dimension in the Feynman gauge. Our final result is  
\begin{eqnarray} 
\lambda_{+,1}(a_c) &=& 
-~ [8\mu^3n^2 + 8\mu^3n + 8\mu^3 + 2\mu^2n^4 + 4\mu^2n^3 - 22\mu^2n^2 
- 24\mu^2n \nonumber \\ 
&&~~~ - 28\mu^2 - 6\mu n^4 - 12\mu n^3 + 14\mu n^2 + 20\mu n + 32\mu 
+ 5n^4 \nonumber \\ 
&&~~~ + 10n^3 + n^2 - 4n - 12]\Gamma(n+2-\mu) 
\Gamma(\mu - 1)\mu C_2(R) \eta_1^{\mbox{o}} \nonumber \\ 
&&~~~~ /[(\mu - 2)^2(n + 2)(n + 1)(n - 1)
\Gamma(2 - \mu)\Gamma(\mu + n)n T(R)] \nonumber \\ && \nonumber \\  
&&+~ \frac{2\mu(\mu - 1)S_1(n) C_2(G) \eta_1^{\mbox{o}}} 
{(2\mu - 1)(\mu - 2)T(R)} \nonumber \\ 
&& \nonumber \\ 
&&-~ [32\mu^5n^2 + 32\mu^5n + 32\mu^5 - 144\mu^4n^2 - 144\mu^4n - 160\mu^4 
- 4\mu^3n^4 \nonumber \\ 
&&~~~ - 8\mu^3n^3 + 240\mu^3n^2 + 244\mu^3n + 316\mu^3 + 16\mu^2n^4 
+ 32\mu^2n^3 \nonumber \\ 
&&~~~ - 180\mu^2n^2 - 196\mu^2n - 306\mu^2 - 20\mu n^4 - 40\mu n^3 + 59\mu n^2
\nonumber \\
&&~~~ + 79\mu n + 146\mu + 8n^4 + 16n^3 - 6n^2 - 14n - 28]\mu C_2(G) 
\eta_1^{\mbox{o}} \nonumber \\ 
&&~~~~ /[8(2\mu - 1)(\mu - 1)^3(\mu - 2)(n + 2)(n + 1)(n - 1)n T(R)]  
\nonumber \\ && \nonumber \\  
&&+~ [32\mu^5n^2 + 32\mu^5n + 32\mu^5 + 8\mu^4n^4 + 16\mu^4n^3 - 120\mu^4n^2 
- 128\mu^4n \nonumber \\
&&~~~ - 160\mu^4 - 32\mu^3n^4 - 64\mu^3n^3 + 160\mu^3n^2 
+ 192\mu^3n + 316\mu^3 + 48\mu^2n^4 \nonumber \\ 
&&~~~ + 96\mu^2n^3 - 78\mu^2n^2 - 126\mu^2n - 306\mu^2 - 31\mu n^4 - 62\mu n^3 
+ 31\mu n \nonumber \\ 
&&~~~ + 146\mu + 7n^4 + 14n^3 + 7n^2 - 28]  
\Gamma(n+2-\mu)\Gamma(\mu - 1)\mu C_2(G) \eta_1^{\mbox{o}} \nonumber \\ 
&&~~~~ /[8(2\mu - 1)(\mu - 1)^2(\mu - 2)
(n + 2)(n + 1)(n - 1)\Gamma(2-\mu)\Gamma(\mu + n)n T(R)] \nonumber \\  
\label{answer} 
\end{eqnarray} 
where we have set $\lambda_+(a_c)$ $=$ $\sum_{i=0}^\infty 
\lambda_{+,i}(a_c)/\Nf^i$. One curious feature of this expression is that not 
all the $\Gamma$-function structures of the individual graphs survive. 
Specifically the terms involving the factor 
$\Gamma(\mu-1)\Gamma(n)/\Gamma(\mu+n)$ which appears in say (\ref{glu5}) have 
cancelled in (\ref{answer}). We can offer no explanation of this and it is not 
clear if it would be a feature of the analogous dimension in other 
(renormalizable) field theories. Moreover, the result is remarkably more 
compact than one would expect given the nature of (\ref{glu5}) and 
(\ref{glu6}).  

It remains to check that (\ref{answer}) is in agreement with perturbation 
theory. Setting $\mu$ $=$ $2$ $-$ $\epsilon$ and Taylor series expanding to 
several orders it is straightforward to produce the exact $n$-dependence for 
the two loop eigenvalue (\ref{eigandim}). Moreover, expanding to three loops we
obtain the following result for the combination of $3$-loop entries of the 
mixing matrix,  
\begin{eqnarray} 
d_{31} + \frac{b_{31} c_1}{d_{11}}  
&=& \frac{64(n^2 + n + 2)^2 (S_1(n))^2 C_2(R)}{3(n + 2)(n + 1)^2(n - 1)n^2 
T(R)} \nonumber \\ 
&& -~ \frac{64(10n^6 + 30n^5 + 109n^4 + 168n^3 + 155n^2 + 76n + 12) S_1(n) 
C_2(R)}{9(n + 2)(n + 1)^3(n - 1)n^3 T(R)} \nonumber \\  
&& -~ 4[33n^{10}\! + 165n^9 - 32n^8 - 1118n^7 - 5807n^6 - 12815n^5 
- 16762n^4 - 13800n^3 \nonumber \\ 
&& ~~~~~ - 7112n^2 - 2112n - 288]C_2(R)/[27(n + 2)(n + 1)^4(n - 1)n^4 T(R)] 
\nonumber \\ 
&& -~ \frac{8(8n^6 + 24n^5 - 19n^4 - 78n^3 - 253n^2 - 210n - 96) S_1(n) 
C_2(G)}{27(n + 2)(n + 1)^2(n - 1)n^2 T(R)} \nonumber \\ 
&& -~ 2[87n^8 + 348n^7 + 848n^6 + 1326n^5 + 2609n^4 + 3414n^3 + 2632n^2 
\nonumber \\ 
&& ~~~~~ + 1088n + 192]C_2(G)/[27(n + 2)(n + 1)^3(n - 1)n^3 T(R)]  
\label{dbcd} 
\end{eqnarray} 
This needs to be compared with the first few non-zero moments of \cite{9}. To
this end we have evaluated (\ref{dbcd}) for even $n$, $2$ $\leq$ $n$ $\leq$ 
$24$, and recorded the values as exact fractions in table 1. The coefficients 
with respect to each colour Casimir have been listed separately. The entries 
with $n$ $\leq$ $8$ are in {\em exact} agreement with the explicit $3$-loop 
$\MSbar$ expressions of \cite{9} at this order in $1/\Nf$. Therefore we believe
(\ref{answer}) is correct. The provision of higher moments in table 1 is to be 
an aid to future checks on the explicit full analytic result as a function of 
$n$ when it becomes available. Ordinarily in the perturbative calculation one 
has an additional check on the result emanating from general considerations. 
For the case when $n$ $=$ $2$ the singlet operators correspond to the energy 
momentum tensor which is a conserved (non-anomalous) physical current. 
Therefore its anomalous dimension vanishes to all orders in perturbation 
theory, \cite{22}. Such a feature ought to be present in the large $\Nf$ 
results and this indeed was apparent in the dimension of the fermionic operator
of \cite{11}. However, in the gluonic case, as can be seen in table 1, the $n$ 
$=$ $2$ entries are non-zero. This is not inconsistent with this principle. If 
one examines the combination of perturbative coefficients which vanishes at $n$
$=$ $2$ one finds it is, at one loop, $(a_1$ $+$ $b_1)$ and $(c_1$ $+$ 
$d_{11}\Nf$ $+$ $d_{12})$ with similar combinations at higher order. However 
the large $\Nf$ combinations are $(a_1$ $-$ $b_1c_1/d_{11})$ and $(d_{12}$ $+$ 
$b_1c_1/d_{11})$ respectively for the fermionic and gluonic operators. 
Therefore clearly the former ought to be zero at $n$ $=$ $2$ but not the 
latter.  

One feature of the form of the mixing matrix (\ref{matdef}) is that the leading
order series in $\Nf$ for $\gamma_{gq}(a)$ depends only on the Casimir 
$C_2(R)$. This is clear from the one and two loop results of \cite{2,3,4,5} as 
well as the three loop results of \cite{9} for $n$ $\leq$ $8$. Therefore if we 
make the assumption that this is true for all $n$ at three loops then we can 
deduce the exact form of the coefficients $d_{31}$ in the $C_2(G)$ sector from 
(\ref{dbcd}). It is given by the last two terms of (\ref{dbcd}) which are each
proportional to $C_2(G)$. Indeed if one examines the second column of table 1, 
one observes that the first four entries are the same as those which appear in 
\cite{9}, after noting that $n_{\! f}$ $=$ $2T(R) \Nf$.  

As another motivation of this study is to provide a window into the structure
of the operator dimension beyond what is currently available, it is a trivial
exercise to expand (\ref{answer}) for example to four loops to deduce 
\begin{eqnarray} 
d_{41} + \frac{b_{41} c_1}{d_{11}}  
&=& \frac{256(n^2 + n + 2)^2 [ S_3(n) + 2 (S_1(n))^3 ] C_2(R)} 
{27(n + 2)(n + 1)^2(n - 1)n^2 T(R)} 
\nonumber \\ 
&& -~ 256 [10n^6 + 30n^5 + 109n^4 + 168n^3 + 155n^2 + 76n + 12 ] \nonumber \\ 
&& ~~~~~~~~ \times (S_1(n))^2 C_2(R)/[27(n + 2)(n + 1)^3(n - 1)n^3 T(R)] 
\nonumber \\ 
&& +~ 256 [ 37n^8 + 148n^7 + 697n^6 + 1573n^5 + 2087n^4 + 1725n^3 + 889n^2
\nonumber \\ 
&& ~~~~~~~~ + 264n + 36 ] S_1(n) C_2(R)/[81(n + 2)(n + 1)^4 (n - 1)n^4 T(R)] 
\nonumber \\ 
&& +~ \frac{256(n^2 + n + 2)^2 \zeta(3) C_2(R)}{9(n + 2)(n + 1)^2(n - 1)n^2 
T(R)} \nonumber \\ 
&& -~ 16 [ 77n^{12} + 462n^{11} + 1481n^{10} + 3170n^9 + 12839n^8 + 37418n^7 
\nonumber \\ 
&& ~~~~~~~ + 75483n^6 + 103718n^5 + 98592n^4 + 64176n^3 + 27672n^2 
\nonumber \\ 
&& ~~~~~~~ + 7200n + 864] C_2(R)/[243(n + 2)(n + 1)^5(n - 1)n^5 T(R)]  
\nonumber \\ 
&& +~ \frac{32(9n^4 + 18n^3 + 79n^2 + 70n + 32) (S_1(n))^2 C_2(G)}
{27(n + 2)(n + 1)^2(n - 1)n^2 T(R)} ~+~ \frac{512 S_1(n) \zeta(3) C_2(G)} 
{27 T(R)} \nonumber \\ 
&& -~ 32 [ 8n^8 + 32n^7 + 95n^6 + 173n^5 + 781n^4 + 1311n^3 + 1176n^2 
\nonumber \\ 
&& ~~~~~~~ + 544n + 96] S_1(n) C_2(G)/[81(n + 2)(n + 1)^3( n - 1)n^3 T(R)] 
\nonumber \\ 
&& -~ \frac{1024(n^2 + n + 1) \zeta(3) C_2(G)}{27(n + 2)(n + 1)(n - 1)n T(R)}  
\nonumber \\ 
&& -~ 8 [ 5n^{10} + 25n^9 + 202n^8 + 658n^7 - 2639n^6 - 10115n^5 - 17896n^4 
- 18216n^3 \nonumber \\ 
&& ~~~~~~ - 11160n^2 - 3840n - 576 ] C_2(G) 
/[243(n + 2)(n + 1)^4(n - 1)n^4 T(R)] \nonumber \\  
\label{4loop} 
\end{eqnarray} 
where $\zeta(n)$ is the Riemann $\zeta$-function. To obtain five and higher 
order combinations of perturbative coefficients at this order in $1/\Nf$ one 
needs to first deduce the corresponding corrections to (\ref{ac}) which are 
encoded in the $O(1/\Nf)$ QCD $\beta$-function, \cite{14}. Again if we make the
assumption that $b_{41}$ involves only $C_2(R)$ then the terms proportional to
$C_2(G)$ in (\ref{4loop}) determine exactly the $C_2(G)$ part of $d_{41}$. 

Finally we make some concluding remarks. First the determination of the 
expression (\ref{answer}) completes the leading order analysis in the large 
$\Nf$ expansion of the unpolarized anomalous dimensions for twist-$2$ 
operators. To complete the full $O(1/\Nf)$ analysis of deep inelastic
scattering one requires information on the corresponding coefficient functions.
This has been achieved for the non-singlet case by various authors, 
\cite{15,16}. As far as we are aware though, the unpolarized singlet case has 
yet to be examined. Second, it ought to be possible to extend our present work 
to determine the anomalous dimensions of the gluonic twist-$2$ singlet 
operators of polarized scattering. We would hope to return to this in a future 
article.  

\vspace{1cm} 
{\bf Acknowledgements.} This work was carried out with the support of 
{\sc PPARC} through an Advanced Fellowship (JAG) and a Postgraduate Studentship
(JFB). 

\vspace{1cm} 

\appendix 

\sect{Basic results.} 

We give here several elementary integration rules for simple chain integrals.
Although they have appeared before (see, for example, \cite{23}), we record the
simplest cases here for completeness and also to fix our notation. First, we 
recall the chain result of \cite{12}, for arbitrary $\alpha$ and $\beta$,  
\begin{equation} 
\int_y \, \frac{1}{(y^2)^\alpha ((x-y)^2)^\beta} ~=~ 
\frac{\nu(\alpha,\beta,2\mu-\alpha-\beta)}{(x^2)^{\alpha+\beta-\mu}} 
\label{cha1} 
\end{equation} 
where $\nu(\alpha,\beta,\gamma)$ $=$ $\pi^\mu a(\alpha) a(\beta) a(\gamma)$ 
and $a(\alpha)$ $=$ $\Gamma(\mu-\alpha)/\Gamma(\alpha)$ which is established
by using Feynman parametrization. The more general moment type chain integral
is given by  
\begin{equation} 
\int_y \, \frac{(\Delta y)^p (\Delta (x-y))^q}{(y^2)^\alpha ((x-y)^2)^\beta} 
{}~=~ \nu_{pq(p+q)}(\alpha,\beta,2\mu-\alpha-\beta+p+q) \frac{(\Delta x)^{p+q}} 
{(x^2)^{\alpha+\beta-\mu}} 
\label{cha2} 
\end{equation} 
where $\nu_{mnp}(\alpha,\beta,\gamma)$ $=$ $\pi^\mu a_m(\alpha) a_n(\beta) 
a_p(\gamma)$, $a_n(\alpha)$ $=$ $\Gamma(\mu-\alpha+n)/\Gamma(\alpha)$ and $m$,
$n$ and $p$ are integers. With (\ref{cha2}) one can, for example, build up 
two loop integrals which appear in the determination of the graph of fig. 3
where the exponent of the top left propagator is replaced by $0$.  

Next we record the explicit expressions for the three loop recurrence relations
we have used, where $\alpha$ and $\beta$ are arbitrary and $n$ is a strictly 
positive integer. Their derivation has been discussed in section 4. 
\begin{eqnarray} 
C_{10\gamma}(\alpha,\beta,n) &=& -~ \frac{2(2\mu+n)(2\mu-\alpha-\beta+n-2)}
{(n+1)(n+2)} C_{10\gamma}(\alpha+1,\beta,n+2) \nonumber \\  
&& -~ \frac{(2\mu-2\alpha+n-3)}{(n+1)} C_{10\gamma}(\alpha+1,\beta,n+1) 
\nonumber \\  
&& +~ [ C^{011}_{00\gamma}(\alpha+1,\beta,n+1) + C_{00\gamma}(\alpha,\beta,n+1) 
- C^{101}_{00\gamma}(\alpha+1,\beta,n+1) ] \nonumber \\  
&& ~~~ \times \frac{(2\mu-\alpha-\beta+n-2)}{(n+1)} \nonumber \\  
&& -~ [ C_{00\gamma}(\alpha+1,\beta,n+2) 
+ C^{011}_{00\gamma}(\alpha+1,\beta,n+2) 
- C^{110}_{00\gamma}(\alpha+1,\beta,n+2) ] \nonumber \\  
&& ~~~ \times \frac{(2\mu+n)(2\mu-\alpha-\beta+n-2)}{(n+1)(n+2)} \nonumber \\  
&& -~ [ C^{021}_{10\gamma}(\alpha+1,\beta-1,n+1) 
- C^{120}_{10\gamma}(\alpha,\beta,n+1) \nonumber \\  
&& ~~~ + C^{021}_{01\gamma}(\alpha+1,\beta-1,n+1) 
- C^{120}_{01\gamma}(\alpha,\beta,n+1) ]/(n+1) 
\end{eqnarray}  
\begin{eqnarray} 
C_{11\gamma}(\alpha,\beta,n) &=& \frac{2(2\mu+n+1)(2\mu-\alpha-\beta+n-2)}
{(n+1)(n+2)} C_{11\gamma}(\alpha+1,\beta,n+2) \nonumber \\  
&& -~ \frac{(2\mu-2\alpha+n-3)}{(n+1)} C_{11\gamma}(\alpha+1,\beta,n+1) 
\nonumber \\  
&& +~ [ 2C^{011}_{00\gamma}(\alpha+1,\beta,n+2)  
- C_{00(\gamma-1)}(\alpha+1,\beta,n+2) ] \nonumber \\  
&& ~~~ \times \frac{(2\mu-\alpha-\beta+n-2)}{(n+1)(n+2)} \nonumber \\  
&& +~ [ C_{10\gamma}(\alpha,\beta,n+1) + C^{011}_{01\gamma}(\alpha+1,\beta,n+1) 
- C^{101}_{01\gamma}(\alpha+1,\beta,n+1) ] \nonumber \\  
&& ~~~ \times \frac{(2\mu-\alpha-\beta+n-2)}{(n+1)} \nonumber \\  
&& -~ [ C_{10\gamma}(\alpha+1,\beta,n+2) 
+ C^{011}_{01\gamma}(\alpha+1,\beta,n+2)  
- C^{110}_{01\gamma}(\alpha+1,\beta,n+2) ] \nonumber \\  
&& ~~~ \times \frac{(2\mu+n+2)(2\mu-\alpha-\beta+n-2)}{(n+1)(n+2)} \nonumber \\ 
&& -~ 2[ C^{021}_{11\gamma}(\alpha+1,\beta-1,n+1) 
- C^{120}_{11\gamma}(\alpha,\beta,n+1) ]/(n+1) 
\end{eqnarray} 
\begin{eqnarray} 
C_{20\gamma}(\alpha,\beta,n) &=& \frac{2(2\mu+n+1)(2\mu-\alpha-\beta+n-2)}
{(n+1)(n+2)} C_{20\gamma}(\alpha+1,\beta,n+2) \nonumber \\  
&& -~ \frac{(2\mu-2\alpha+n-3)}{(n+1)} C_{20\gamma}(\alpha+1,\beta,n+1) 
\nonumber \\  
&& +~ \frac{2(2\mu-\alpha-\beta+n-2)}{(n+1)(n+2)}
C^{011}_{00\gamma}(\alpha+1,\beta,n+2) \nonumber \\  
&& +~ [ C^{011}_{10\gamma}(\alpha+1,\beta,n+1) 
+ C_{10\gamma}(\alpha,\beta,n+1) 
- C^{101}_{10\gamma}(\alpha+1,\beta,n+1) ] \nonumber \\  
&& ~~~ \times \frac{(2\mu-\alpha-\beta+n-2)}{(n+1)} \nonumber \\  
&& -~ [ C_{10\gamma}(\alpha+1,\beta,n+2) 
+ C^{011}_{10\gamma}(\alpha+1,\beta,n+2)  
- C^{110}_{10\gamma}(\alpha+1,\beta,n+2) ] \nonumber \\  
&& ~~~ \times \frac{(2\mu+n+2)(2\mu-\alpha-\beta+n-2)}{(n+1)(n+2)} \nonumber \\ 
&& -~ [ C^{021}_{20\gamma}(\alpha+1,\beta-1,n+1) 
- C^{120}_{20\gamma}(\alpha,\beta,n+1) \nonumber \\ 
&& ~~~ + C^{021}_{02\gamma}(\alpha+1,\beta-1,n+1) 
- C^{120}_{02\gamma}(\alpha,\beta,n+1) ]/(n+1) 
\end{eqnarray} 
We note that in their application to the integrals we require in this paper the
first two terms on the right side which are the same function on the left but 
with $\alpha$ replaced by $\alpha$ $+$ $1$ will be zero by the dimension 
argument. 

\sect{Explicit check for $C_{201}(2\mu-4+\Delta,\mu-1+\Delta,1)$.} 

Throughout our discussion we have mentioned carrying out explicit checks on 
recurrence relations for low values of $n$. We give a detailed illustration of 
this here by summarizing the calculation of the result for 
$C_{201}(2\mu-4+\Delta,\mu-1+\Delta,1)$. Unlike the procedure used to construct
recurrence relations we need to compute all of the invariant amplitudes which
arise in the tensor decomposition of the corresponding integral without the
null vector $\Delta_\mu$. At the end of this calculation contracting with the
appropriate number of these vectors will give us a $\mu$-dependent result
which will agree with the general $n$-dependent expression evaluated at $n$ 
$=$ $1$. Therefore if we label the indices of the $y$-propagator of fig. 7 as
$\mu$ and $\nu$ and that of the $u$-propagator as $\sigma$, then the integral
containing $C_{201}(2\mu-4+\Delta,\mu-1+\Delta,1)$ decomposes into  
\begin{equation} 
A \eta_{\mu\nu}x_\sigma ~+~ B [ \eta_{\mu\sigma} x_\nu + \eta_{\nu\sigma} 
x_\mu ] ~+~ C \frac{x_\mu x_\nu x_\sigma}{x^2} 
\end{equation} 
The three contractions $\eta^{\mu\nu}x^\sigma$, $\eta^{\mu\sigma}x^\nu$ and 
$x^\mu x^\nu x^\sigma$ yield the respective combinations $(2\mu A$ $+$ $2B$ $+$
$C)$, $(A$ $+$ $(2\mu+1)B$ $+$ $C)$ and $(A$ $+$ $2B$ $+$ $C)$. Using identities
like $xy$ $=$ $\half[x^2$ $+$ $y^2$ $-$ $(x-y)^2]$ we can rewrite the resulting
integrals as a sum of scalar integrals. Again the majority of these are 
computable using the simple rules analogous to that of fig. 5. However, several
basic results are necessary which it turns out have already been computed in
the calculation of \cite{14} and we merely quote these here  
\begin{eqnarray} 
C_{001}(2\mu-4+\Delta,\mu-1+\Delta,0) &=& - ~ \frac{(\mu-1)^2\nu^2(1,1,2\mu-2)} 
{\Gamma(\mu)\Delta} \nonumber \\ 
C_{001}(2\mu-5+\Delta,\mu-1+\Delta,0) &=& - ~ \frac{(\mu-1)^2(2\mu^2-9\mu+16)
\nu^2(1,1,2\mu-2)}{4 \Gamma(\mu+1)\Delta} \\ 
C_{001}(2\mu-4+\Delta,\mu-2+\Delta,0) &=& - ~ \frac{(\mu-1)^2(2\mu^2-5\mu+4)
\nu^2(1,1,2\mu-2)}{4 \Gamma(\mu+1)\Delta} \nonumber  
\end{eqnarray} 
Therefore the above combinations become 
\begin{eqnarray} 
2\mu A + 2B + C &=& -~ \frac{(2\mu - 7)(\mu - 1)(\mu - 2)(\mu - 3) 
\nu^2(1,1,2\mu-2)}{6\Gamma(\mu+2)\Delta} \nonumber \\  
A + (2\mu+1)B + C &=& -~ \frac{(2\mu^4 + 11\mu^3 - 121\mu^2 + 374\mu - 324) 
(\mu - 1)\nu^2(1,1,2\mu-2)}{48\Gamma(\mu + 2)\Delta} \\  
A + 2B + C &=& -~ \frac{(6\mu^4 + 53\mu^3 - 333\mu^2 + 1408\mu - 1464) 
(\mu - 1)\nu^2(1,1,2\mu-2)}{96\Gamma(\mu + 3)\Delta} \nonumber  
\end{eqnarray} 
Finally, solving for $A$, $B$ and $C$ the uncontracted integral is 
\begin{eqnarray} 
C_{201}(2\mu-4+\Delta,\mu-1+\Delta,1) &=& 
-~ \left[ \eta_{\mu\nu}x_\sigma (26\mu^4 - 261\mu^3 + 541\mu^2 - 576\mu
+ 120) \right. \nonumber \\ 
&& \left. ~~~~ + (4\mu^4 + 36\mu^3 - 143\mu^2 + 168\mu - 56)(\mu-3) \right. 
\nonumber \\ 
&& \left. ~~~~~ \times (\eta_{\mu\sigma}x_\nu + \eta_{\nu\sigma}x_\mu) 
\right. \nonumber \\  
&& \left. ~~~~ + 2(2\mu^5 + 13\mu^4 + 22\mu^3 + 707\mu^2 - 1320\mu + 504) 
\right. \nonumber \\ 
&& \left. ~~~~~~ \times \frac{x_\mu x_\nu x_\sigma}{x^2} \right] (\mu-1)
\nu(1,1,2\mu-2) \nonumber \\  
&& ~~~~~ /[96(2\mu-1)\Gamma(\mu+3) \Delta]  
\end{eqnarray}  
whence the check on the recurrence relation we require is given by the 
coefficient of the $x^\mu x^\nu x^\sigma$ term. It is in agreement with the 
value of (\ref{c20141n}) at $n$ $=$ $1$.

\newpage 
{\begin{table} 
\begin{center} 
\begin{tabular}{c||r|r} 
$n$ & $C_2(R)/T(R)$ coefficient & $C_2(G)/T(R)$ coefficient \\ 
\hline  
& & \\ 
2 & $ - \, \frac{2716}{243}$ & $ - \, \frac{4232}{243}$ \\ 
& & \\ 
4 & $ - \, \frac{3765671}{607500}$ & $ - \, \frac{757861}{60750}$ \\ 
& & \\ 
6 & $ - \, \frac{373918478}{72930375}$ & $ - \, \frac{26390948}{2083725}$ \\ 
& & \\ 
8 & $ - \, \frac{261337619387}{54010152000}$ & 
$ - \, \frac{420970849}{32148900}$ \\ 
& & \\ 
10 & $ - \, \frac{1994718278948}{420260754375}$ &
$ - \, \frac{2752314359}{203762790}$ \\ 
& & \\ 
12 & $ - \, \frac{26908165162400891}{5703275664286200}$ &
$ - \, \frac{2635361358193}{189919269540}$ \\ 
& & \\ 
14 & $ - \, \frac{16449437567815379}{3489766577797500}$ &
$ - \, \frac{4616790551}{325061100}$ \\ 
& & \\ 
16 & $ - \, \frac{331641918853126821883}{70281957317539968000}$ &
$ - \, \frac{35631283224283}{2458394452800}$ \\ 
& & \\ 
18 & $ - \, \frac{7462146988179117961999}{1578449269228939482375}$ &
$ - \, \frac{249465132233659}{16907213454300}$ \\ 
& & \\ 
20 & $ - \, \frac{8216273557478999257489}{1734325382898288180000}$ &
$ - \, \frac{1340260211455733}{89401072761000}$ \\ 
& & \\ 
22 & $ - \, \frac{228675683353467573965719}{48168621150965760797352}$ &
$ - \, \frac{648057715871521}{42614381628819}$ \\ 
& & \\ 
24 & $ - \, \frac{45507171465506238473037583}{9566587391661149850000000}$ &
$ - \, \frac{53358759817454947}{3463516806750000}$ \\ 
& & \\ 
\end{tabular} 
\end{center} 
\end{table} 
\begin{center} 
{Table 1. Coefficients of $\left[ d_{31} \right.$ $+$ $\left. 
b_{31}c_1/d_{11} \right]$ as a function of moment.} 
\end{center} } 

\newpage 
{\epsfysize=2.5cm 
\epsfbox{qcddisgraph1.ps}}  
\vspace{1cm} 
{\bf Fig. 1. Operator insertion in gluon $2$-point function.} 

\vspace{2.5cm} 
{\epsfysize=7cm 
\epsfbox{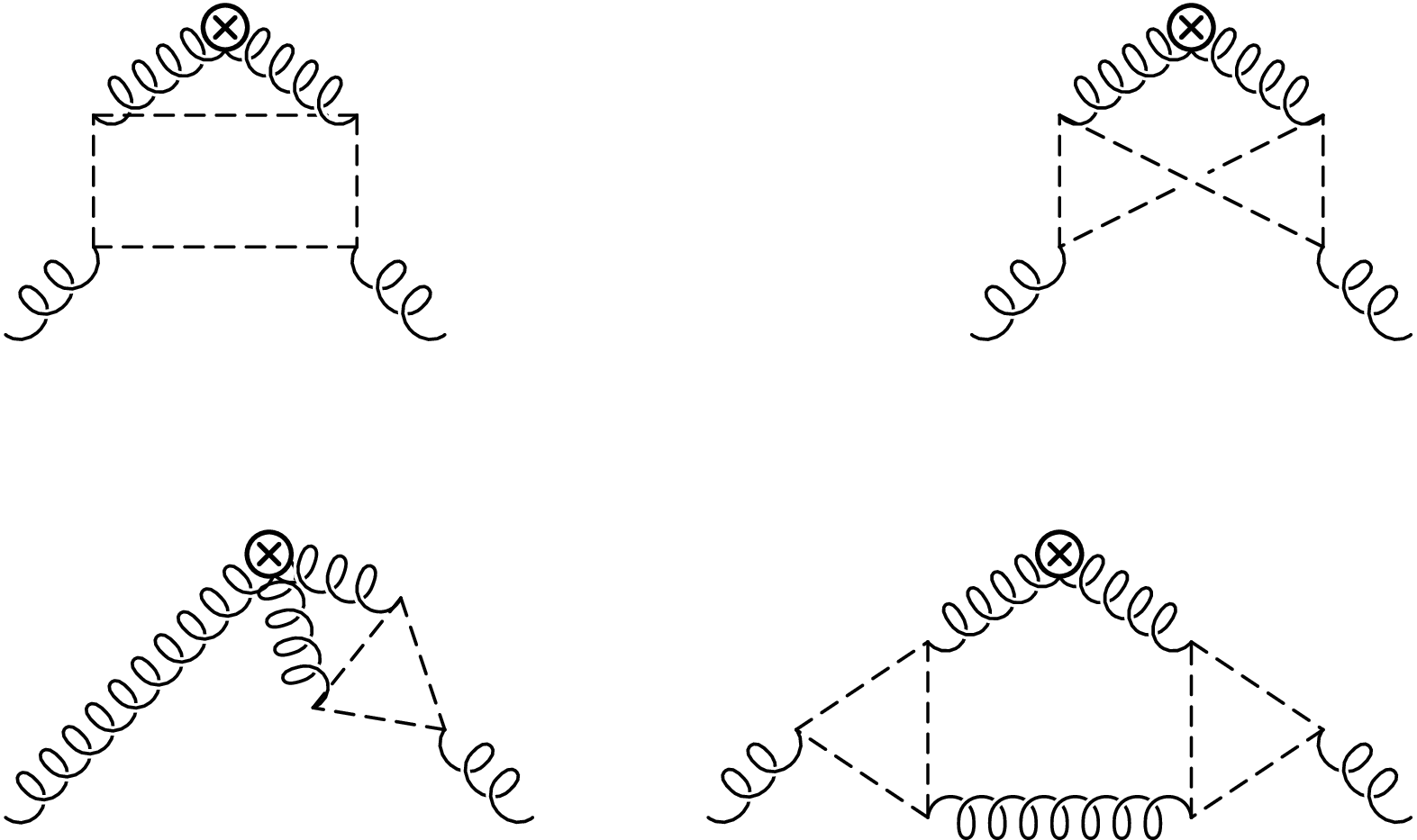}}  
\vspace{1cm} 
{\bf Fig. 2. Leading order diagrams for $\lambda_+(a_c)$.} 

\vspace{2.5cm} 
{\epsfysize=3cm 
\epsfbox{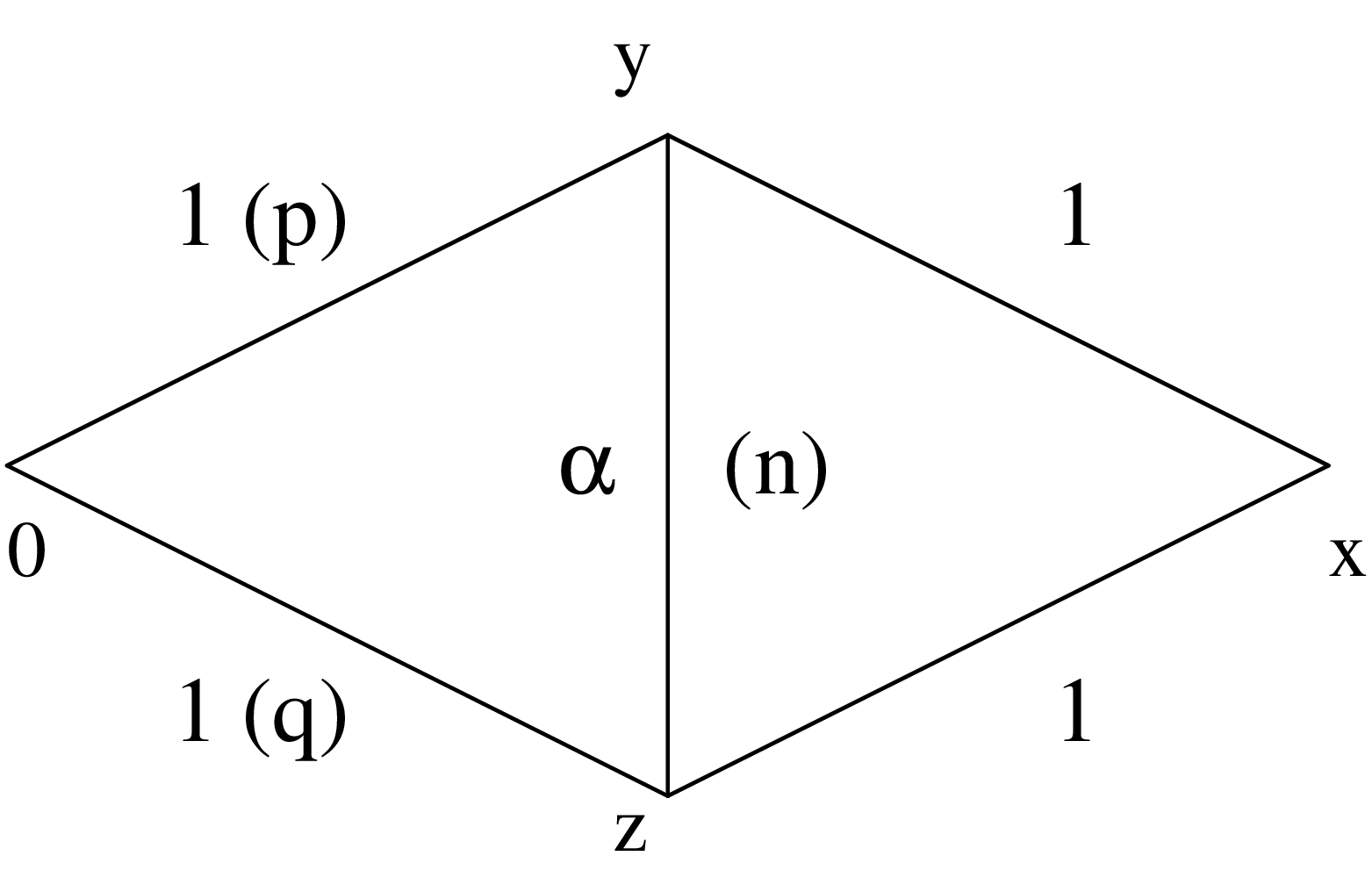}}  
\vspace{1cm} 
{\bf Fig. 3. Definition of $G_{pqn}(\alpha)$.} 

\newpage 
{\epsfysize=3cm 
\epsfbox{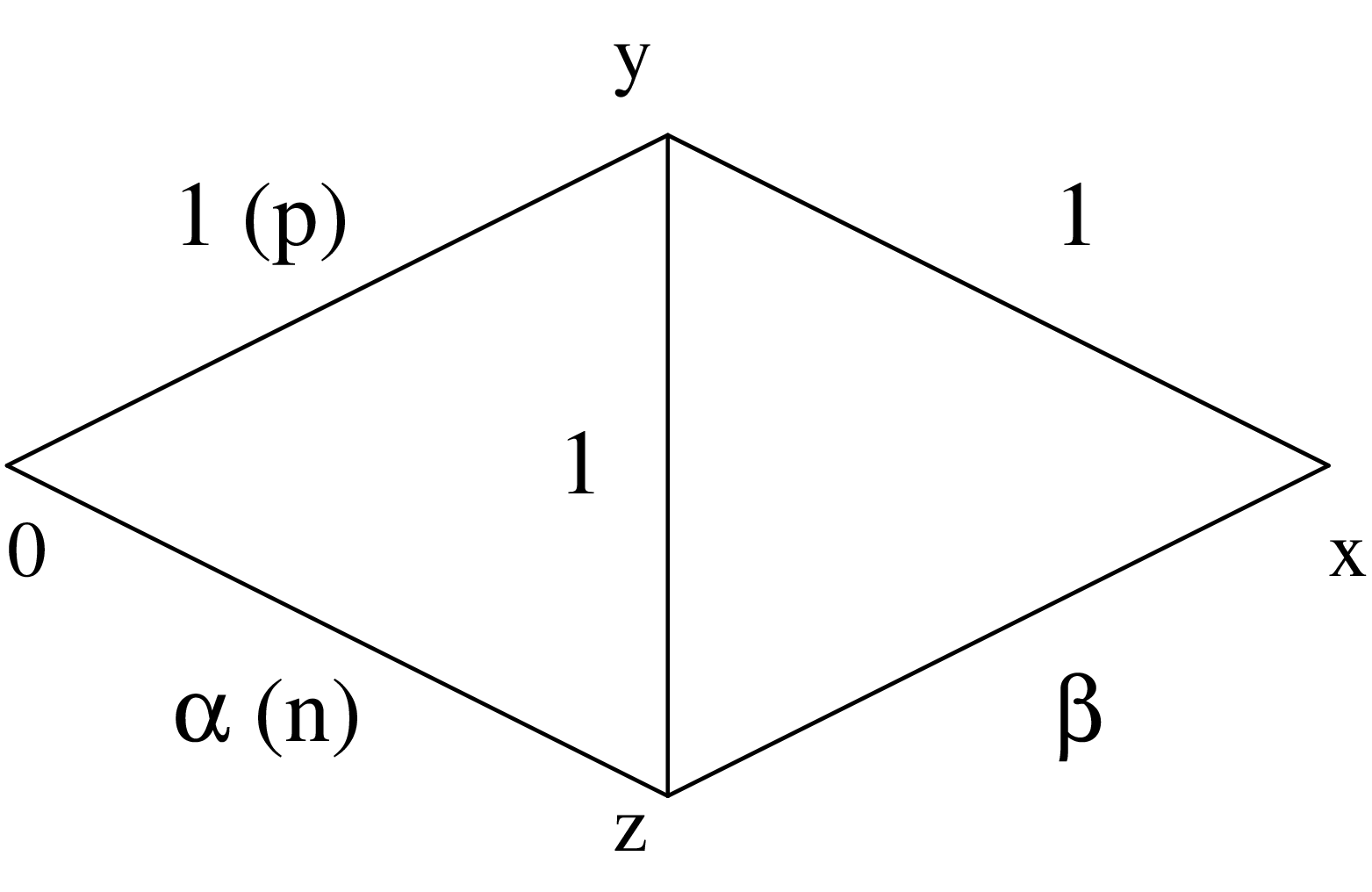}}  
\vspace{0.5cm} 
{\bf Fig. 4. Definition of $H_{pn}(\alpha,\beta)$.} 

\vspace{1.3cm} 
{\epsfysize=5.5cm 
\epsfbox{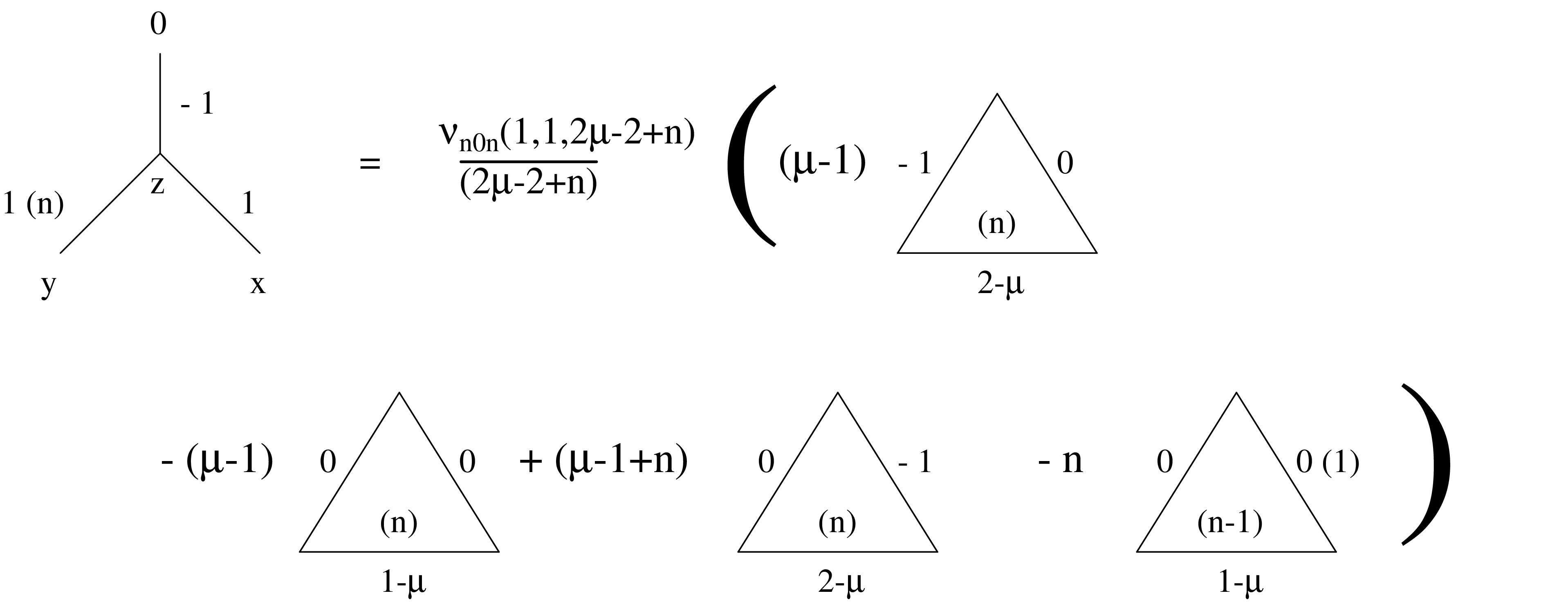}}  
\vspace{0.5cm} 
{\bf Fig. 5. Example of elementary integral rule.} 

\vspace{1.3cm} 
{\epsfysize=3cm 
\epsfbox{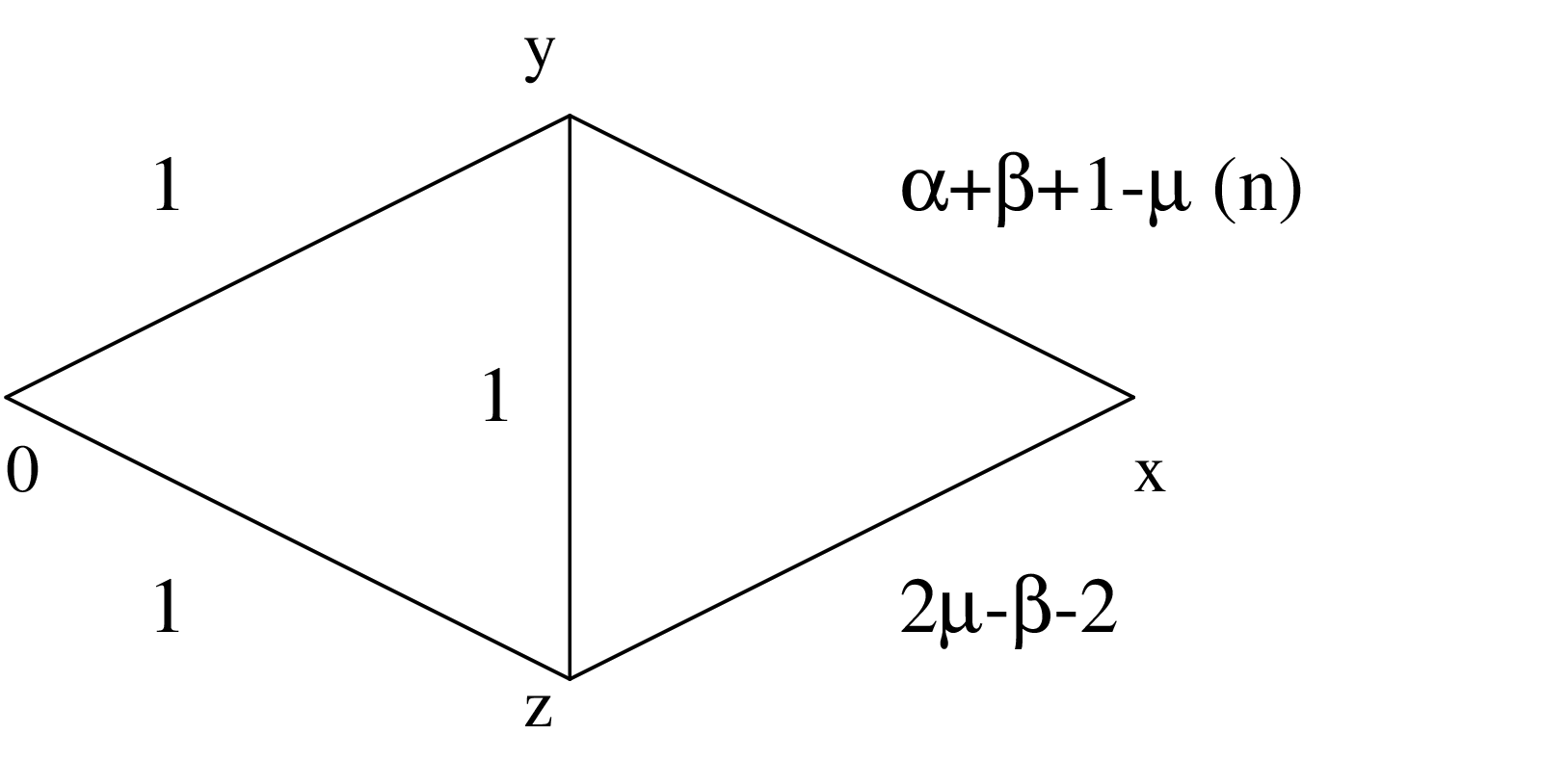}}  
\vspace{0.5cm} 
{\bf Fig. 6. Intermediate integral in calculation of $H_{pn}(\alpha,\beta)$.} 

\vspace{1.3cm} 
{\epsfysize=3cm 
\epsfbox{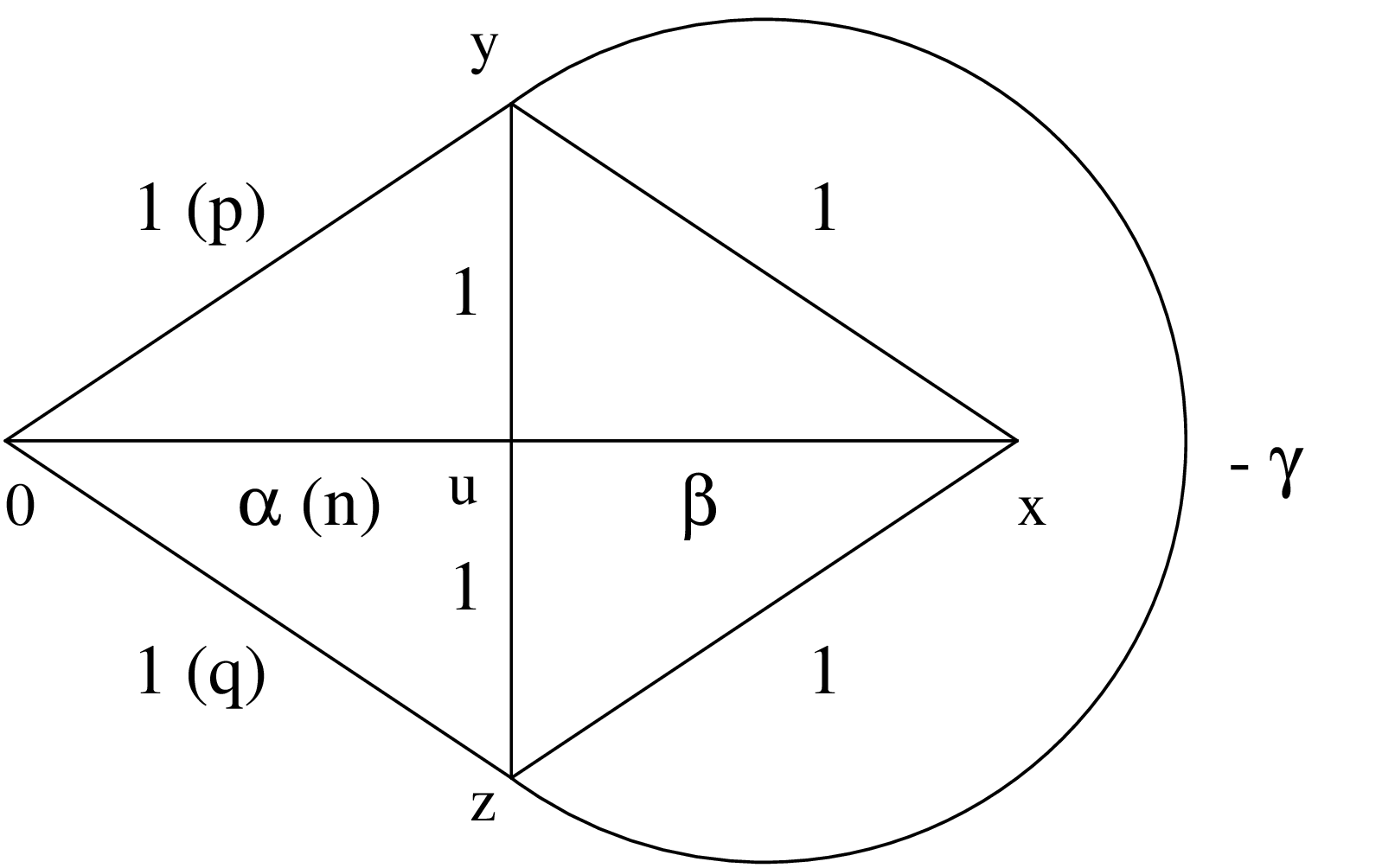}}  
\vspace{0.5cm} 
{\bf Fig. 7. Definition of $C_{pq\gamma}(\alpha,\beta,n)$.} 

\end{document}